\documentclass[11pt]{article}
\usepackage[margin=1in]{geometry}
\usepackage[T1]{fontenc}
\usepackage{lmodern,microtype,amsmath,amssymb,booktabs,array,graphicx,xcolor,enumitem}
\usepackage{hyperref}
\usepackage{caption}
\usepackage{placeins}
\hypersetup{colorlinks=true,linkcolor=teal,citecolor=teal,urlcolor=teal,pdftitle={When Is Fresh AI Stock Analysis Worth It?},pdfauthor={Russlan Ramdowar},pdfsubject={A retrospective study of research refresh budgets in a multi-agent investment platform},pdfkeywords={multi-agent systems, financial forecasting, research budgets, value of information, forecast revision, iPulse AI}}
\setlist{nosep,leftmargin=*}

\title{\textbf{When Is Fresh AI Stock Analysis Worth It?}\\[7pt]\large Allocating a Limited Research Budget Across Price Updates,\\Deep Forecast Refreshes, and Editorial Reviews of Top Stock Picks}
\author{Russlan Ramdowar\\Future Edge Group FZE, United Arab Emirates\\\href{mailto:russlan@ipulseai.com}{russlan@ipulseai.com}\\\small ORCID: \href{https://orcid.org/0009-0004-9311-6957}{0009-0004-9311-6957}}
\date{4 November 2026 --- Version 2.1.0}

\begin{document}
\maketitle
\begin{abstract}
When is another AI research run worth more than keeping that money invested? For long-horizon investment platforms covering hundreds of equities, fresh reasoning must improve decisions enough to cover its recurring cost. This resource-allocation problem underpins iPulse AI. Using an archive of 8,120 production reports on a balanced panel of 203 stocks across four production batches (retaining ten matched agent configurations across seven investment frameworks), we measure how much decision value can be retained by updating current prices against existing forecasts before paying for fresh multi-agent inference. Generating these reports consumed nearly 100 million output and reasoning tokens. Across three refresh transitions, we compare (1) old price targets repriced at current market quotes (with and without a volatility-scaled shock safeguard), (2) a zero-inference mechanical reset that re-anchors old predicted growth rates to current prices, and (3) fresh deep reports.   In our primary economic evaluation---selecting equal-weight Top 10 USD portfolios from one-year forecasts net of 10 bps transaction costs and a USD 3,000 all-in research charge on a USD 300,000 account---shock-guarded old targets beat SPY across all three short (3-month) holding windows without incurring new inference costs, averaging 2.25\% versus 1.66\% for SPY and 4.24\% for fresh reports in unchanged-model runs. The thesis half-life question becomes visible in the longer window: extending the first holding window to nearly six months without a refresh causes the guarded reuse portfolio to fall to -11.26\%, whereas fresh reports return +8.55\% (SPY: 3.46\%). Supporting 90-day forecast diagnostics reveal why inexpensive baselines compete so effectively over shorter windows: mechanically resetting old forecast paths to current prices eliminates 98\% of the Mean Absolute Error (MAE) gap between stale targets (15.26 pp) and fresh LLM analysis (13.51 pp vs. 13.55 pp for reset), yet still turns over 60\%--90\% of the Top 10 shortlist. Expanding the shortlist to Top 30 reverses the unchanged-model comparison in favor of guarded reuse (4.17\% vs. 1.37\% for fresh), and leave-one-out stress tests highlight strong dependence on individual outlier winners (MRNA). A separate audit of nine production editions shows substantial changes in published Top 10 membership between editions, already present in their incoming candidate lists. Within each edition, the final editorial screen retained its incoming list without replacements. Finally, a USD 300,000 account funding USD 3,000 full-universe runs shows why frequency matters: monthly budgets approach the realized short-window value of refreshes, while weekly budgets require substantial additional gains.
\end{abstract}

\section{The Budget Question Behind Fresh Analysis}

Imagine maintaining fundamental research on 203 stocks while investing in only ten of them at a time. Market prices fluctuate every day, but a company's underlying business trajectory rarely changes overnight. Under a finite compute budget, the central operational question is how much decision value can be retained simply by updating daily market prices against existing research, and when new evidence warrants paying for a fresh multi-agent deep dive.

A falling share price can signal a better bargain or a broken investment thesis. Updating implied upside is inexpensive; deciding whether the original target remains defensible requires evidence. The practical value of a fresh run lies in resolving that distinction before capital is committed.

Long-horizon business research is designed to remain useful well beyond a single day's price movements. As Warren Buffett framed it in his 1996 shareholder letter, sound equity ownership rests on evaluating businesses over years rather than reacting to every quote \cite{buffett1996}. That principle is the very reason we invented iPulse AI, an open agentic investment research platform developed by Future Edge Group FZE. Rather than acting as a high-frequency trading bot, iPulse AI separates continuous, inexpensive price monitoring from episodic, compute-intensive thesis rebuilding---reserving deep multi-agent inference for scheduled batch refreshes or material changes in a company's business, financing, or operating environment.

This resource-allocation challenge applies to any organization that monitors a broad candidate universe but acts on a small shortlist, including institutional investment platforms, enterprise risk monitors, procurement systems, and automated due-diligence services. Even as per-token API prices decline, running more frequent inference is not automatically beneficial: regenerating multi-agent reports when fundamentals have not changed can repeat existing context, inject stochastic noise, and churn portfolio rankings without improving decisions.

\begin{figure}[!htbp]\centering
\includegraphics[width=\linewidth]{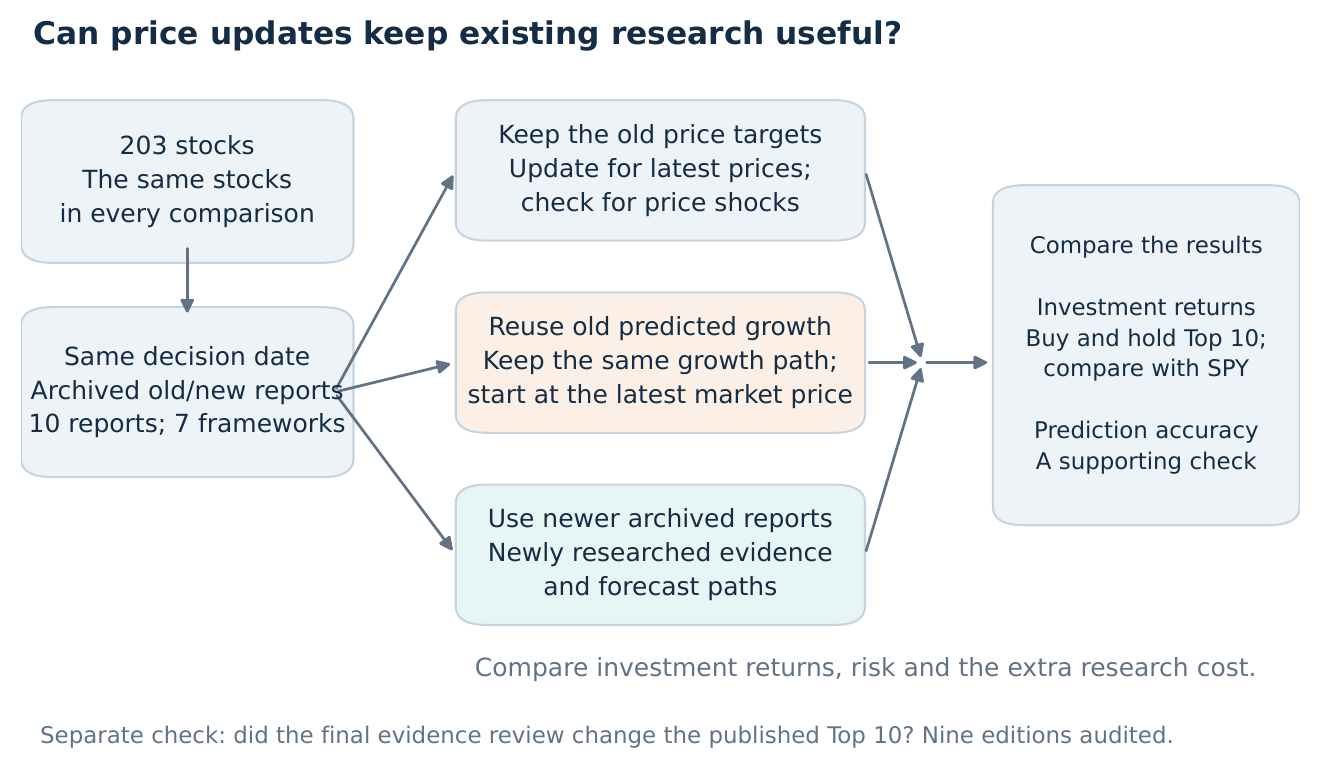}
\caption{Experimental overview. Across a fixed panel of 203 stocks, we compare price-updated old targets (with and without a price-shock safeguard), mechanically reset growth paths, and fresh multi-agent reports. Our primary economic test evaluates one-year signals in Top 10 USD buy-and-hold portfolios against SPY net of modeled trading and compute costs (with Top 30 as a breadth sensitivity). Completed-quarter forecast accuracy and a nine-edition editorial audit provide supporting evidence.}\label{fig:workflow}
\end{figure}

To analyze this trade-off rigorously, we distinguish the update policies and editorial screening that are often conflated under the single word update (Figure~\ref{fig:workflow}):

\textbf{A Price Update (Fixed-Target Repricing):} Recalculating the remaining valuation upside of an existing price target using the latest market close, paired in the evaluated historical implementation with a volatility-scaled shock safeguard.

\textbf{A Mechanical Path Reset (Shape Reset):} Keeping the old report's predicted relative growth path over the remaining horizon, but re-anchoring that trajectory to today's market price---an inference-free baseline that shifts the implied price target without new LLM reasoning.

\textbf{A Deep Refresh (Fresh Reports):} Running the multi-agent council on updated market data and fundamentals to generate new investment theses and numerical forecast paths.

\textbf{An Editorial Screen:} Checking whether dated news or corporate developments invalidate the thesis behind a shortlisted stock prior to publication.

\subsection{Contributions of This Study}

We evaluate the useful life of retained AI equity research through a paired archival comparison and an explicit compute-budget framework. Our study makes five primary contributions:

\textbf{Economic Evaluation of Research Reuse:} A buy-and-hold portfolio comparison against SPY using one-year forecasts net of modeled trading fees and LLM compute charges, showing successful short-window examples and an unfavorable nearly six-month extension.

\textbf{Decomposition of Level Drift vs. Trajectory Revision:} A paired forecast accuracy comparison that separates fixed-target repricing from an inference-free mechanical reset of the old forecast shape, demonstrating that re-anchoring to the current stock price accounts for 98\% of the MAE reduction achieved by fresh LLM reports.

\textbf{Isolation of Model Stability vs. System Evolution:} A balanced 203-stock panel evaluated both within unchanged LLM release identifiers and across an evolving production platform.

\textbf{The Accuracy--Selection Paradox:} Empirical evidence that virtually identical average forecast error ($\Delta\text{MAE} = -0.04\text{ pp}$) can coexist with 60\%--90\% turnover in Top 10 shortlists and opposing portfolio returns at Top 10 versus Top 30 breadth.

\textbf{Production Editorial Audit \& Compute Economics:} An audit of nine published stock-pick editions alongside measured report-token workloads and a capital-scaled break-even model for allocating a limited research budget.

\subsection{Relationship to Prior Work on iPulse AI}

All forecasts evaluated here come from historical iPulse AI production runs generated prior to market outcomes. Earlier studies on this platform examined how inter-agent agreement relates to forecast risk versus reliability \cite{ramdowar-agree} and how disclosing model identity shapes multi-agent synthesis \cite{ramdowar-models}. The present paper addresses a distinct operational question: the marginal decision value of refreshing multi-agent analysis over time versus reusing existing theses with daily price updates.

\section{Related Work}

Cost-aware LLM inference is an active area of systems research. FrugalGPT \cite{frugalgpt} demonstrates how query routing and model cascades reduce API spend for incoming prompts. In our setting, the core decision is not merely which model to call for a new query, but when and whether to rerun an existing multi-agent council at all when a zero-inference price update is available.

Dynamic forecasting benchmarks such as ForecastBench \cite{forecastbench} emphasize evaluating predictions on events whose outcomes are unknown at submission time. While our study analyzes an existing production archive retrospectively rather than as a preregistered prospective trial, every forecast path in our panel was generated and timestamped in production before the corresponding market entry and realization dates. Recent work on LLM-assisted time-series revision \cite{lastmile} and retrieval-based event forecasting \cite{simplicity} similarly highlights the need to weigh agentic complexity against token expenditure. We extend this literature by decomposing forecast updates into fixed-target repricing, shock-guarded repricing, mechanical path resetting, and full multi-agent regeneration.

Finally, operations research has long distinguished statistical prediction error from downstream decision quality. The Smart "Predict, then Optimize" (SPO) framework \cite{spo} incorporates decision objectives directly into model training, while Bailey et al. \cite{backtest} show how easily backtest selection can overfit historical noise. Building on these insights, we evaluate forecast reuse across both statistical loss (MAE/RMSE) and downstream portfolio selection (Top 10 and Top 30), reporting all sensitivity cuts transparently so that sensitivity to portfolio breadth and outlier winners is fully visible.

\section{Data: Tracking an Evolving Production Platform}

\subsection{Coverage, Agent Architecture, and the 203-Stock Balanced Panel}

The evaluation archive was frozen on 2 November 2026 with market observations through 30 October 2026 (archive SHA-256 recorded in Appendix F). Across production Batches 3--7, the platform covered 427 distinct assets, including 384 equities (with 383 equities in Batch 7; Table~\ref{tab:coverage}).

\begin{table}[!htbp]\centering\small
\caption{Historical production coverage across Batches 3--7 prior to balanced-panel matching.}\label{tab:coverage}
\resizebox{\linewidth}{!}{\begin{tabular}{lrrrl}\toprule
Batch & All assets & Stocks & Reports & Role here\\\midrule
3 &247&218&2,718&Old paths for first transition\\
4 &312&283&3,434&First fresh-quarter comparison\\
5 &312&283&3,736&Evolving-system comparison\\
6 &377&348&4,523&Second same-model comparison\\
7 &426&383&5,964&Editorial audit; quarter pending\\\bottomrule
\end{tabular}}
\end{table}

In its current production architecture (reflected in Batch 7), iPulse AI operates 14 opinion configurations across nine investment frameworks, including model and execution-mode variants, topped by a consensus synthesis layer. In the matched historical intersection (Batches 3--6), four of seven frameworks have a single mode while three have both modes, yielding 10 matched configurations with continuous coverage across all four batches (listed in Appendix A). To ensure strict apples-to-apples comparability without conflating changes in agent count with forecast refresh effects, our paired historical accuracy and portfolio tests evaluate the exact 10-configuration intersection across Batches 3--6, while our compute-budget analysis in Section 7 measures the 10-configuration historical panel.

Batches 1--2 are excluded because their early prompt schemas and output protocols differed structurally from later runs. Intersecting complete-quarter coverage across Batches 3--6 yields 206 equities, and excluding three split-sensitive listings leaves our balanced panel of 203 stocks. Retaining the exact same 203 stocks across every transition fixes the comparison universe.

Across the 203 stocks, the two same-model transitions ($3\to4$ and $5\to6$) provide $203 \times 2 = 406$ stock-transition councils (4,060 paired report comparisons), while the evolving-system series ($3\to4$, $4\to5$, and $5\to6$) provides $203 \times 3 = 609$ councils (6,090 paired report comparisons) drawn from $203 \times 10 \times 4 = 8,120$ unique archived reports.

\subsection{Same Model Release vs. Evolving Platform Capabilities}

We track exact archived model release identifiers across batches:

Batches 3 and 4 ran on 20251119\_preview\_release (Gemini 3 Pro Preview).

Batches 5 and 6 ran on 20260201\_preview\_release (Gemini 3.1 Pro Preview).

Accordingly, transitions $3\to4$ and $5\to6$ hold the underlying LLM release fixed (same-model series), whereas $4\to5$ captures a combined model-and-platform upgrade (evolving-system series). Alongside market evolution, several platform enhancements occurred across these batches:

\textbf{Batch 3:} Introduced standardized global macroeconomic context and output schema v5.

\textbf{Batch 4:} Added instruction repetition and long-context hardening (schema v5 with a small v6 cohort).

\textbf{Batch 5:} Upgraded the base model release and introduced structured equity fundamentals, a dedicated consensus narrative workflow, and compact schema v9.

\textbf{Batch 6:} Added currency-normalized fundamentals and expanded asset coverage, followed by the governed editorial review workflow.

To isolate forecast revision from changes in aggregation logic, we aggregate the 10 matched configurations using a fixed two-step rule: we first average execution modes within each framework, and then assign equal weight ($1/7$) across the seven analytical frameworks (Equation 7 in Appendix A).

\section{Methodology: Four Update Policies and Two Evaluation Lenses}

\subsection{Decision Clocks, Entry Prices, and Forecast Horizons}

For each stock and batch transition, the decision date $t$ is aligned to the first saved market close after the newest report in that stock's batch cohort is generated (85--89 calendar days prior to the nominal first-quarter target $T$). This ensures that no forecast is evaluated from a price preceding cohort completion. All comparison arms for a given stock and transition share the exact same actual entry price $P(t)$, target date $T$, and realized market price $P(s_T)$, where $s_T \le T$ is the last available market close on or before nominal expiry.

Let $P_o$ denote the old report's anchor price and $G_o(s)$ its predicted cumulative wealth factor at date $s$ (interpolated geometrically between quarterly knots, with $G_o = 1$ at the anchor). The old report's implied absolute target price at $T$ is $P_o G_o(T)$. Similarly, let $P_n G_n(T)$ denote the fresh report's implied target price. Expressed as percentage returns from the common entry price $P(t)$, the three foundational per-report forecasts are:

\begin{equation}
\widehat{r}^{\mathrm{fixed}}_{t,T} = 100\left[\frac{P_o G_o(T)}{P(t)} - 1\right], 
\end{equation}

\begin{equation}
\widehat{r}^{\mathrm{reset}}_{t,T} = 100\left[\frac{G_o(T)}{G_o(t)} - 1\right], 
\end{equation}

\begin{equation}
\widehat{r}^{\mathrm{fresh}}_{t,T} = 100\left[\frac{P_n G_n(T)}{P(t)} - 1\right]. 
\end{equation}
The realized price return over the interval is $r_{t,T} = 100[P(s_T)/P(t) - 1]$.

\subsection{Why Fixed Targets, Shock Guards, and Mechanical Resets Must Be Separated}

\begin{figure}[!htbp]\centering
\includegraphics[width=\linewidth]{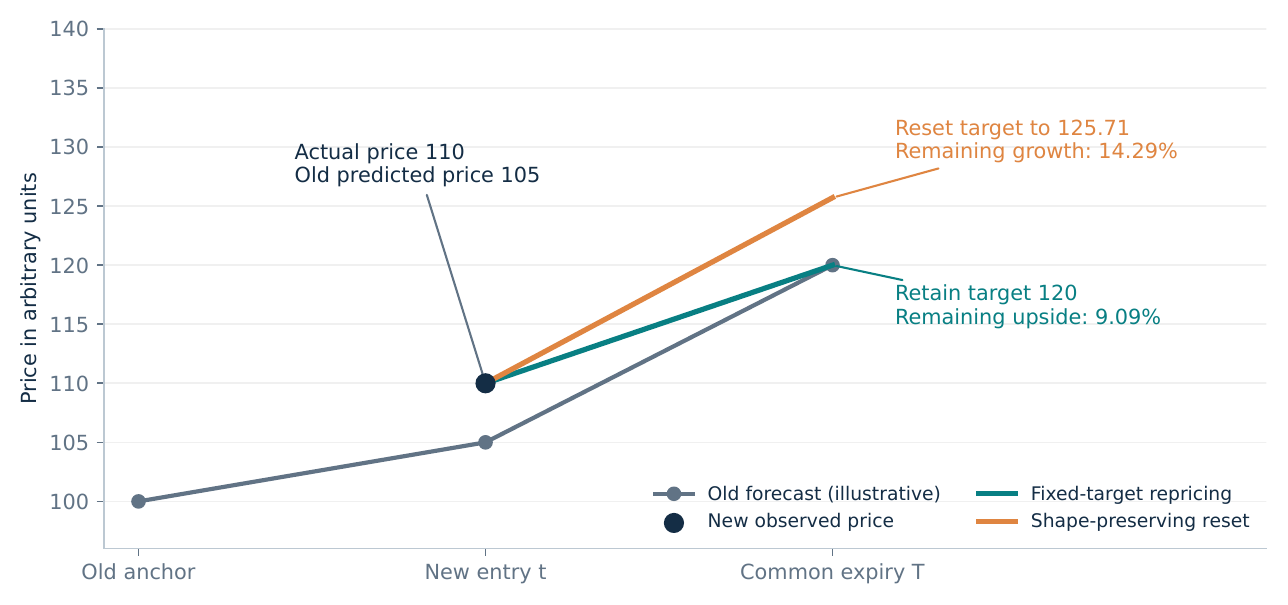}
\caption{Mechanics of forecast reuse. When the observed price at entry $t$ (110) exceeds the old path's interim prediction (105), fixed-target repricing retains the 120 target and compresses remaining upside to 9.09\%, whereas a shape-preserving reset shifts the implied terminal target to 125.71 (+14.29\%). A fresh report can revise both the target level and the trajectory using updated evidence.}\label{fig:baselines}
\end{figure}

When a stock's market price diverges from an earlier forecast path, an automated system can handle that divergence in three distinct ways without generating a new LLM report (Figure~\ref{fig:baselines}):

Unguarded Fixed-Target Repricing ($\widehat{r}^{\mathrm{fixed}}$): Retains the original absolute price target $P_o G_o(T)$ and recalculates remaining upside from today's quote $P(t)$. If the stock has risen, remaining upside shrinks; if the stock has fallen, remaining upside expands.

Shock-Guarded Repricing ($\widehat{r}^{\mathrm{guarded}}$, Historical D3 Rule Evaluated Here): In live portfolio selection, naive fixed-target repricing creates a classic value trap: if a stock drops 40\% because its business thesis broke, dividing the old target by the crashed price mechanically promotes the stock to the top of the buy list. To prevent falling knives from dominating rankings between deep refreshes, the evaluated historical daily recomputation engine (D3) applied a non-linear shock dampener when the current price $P(t)$ deviates sharply from the old interpolated reference trajectory $P_{\mathrm{ref}}(t) = P_o [G_o(T)]^{\tau}$, where $\tau = \min(1, (t - t_o)/(T - t_o))$ is the elapsed fraction of the forecast horizon. Let $d=t-t_o>0$, $H=T_o-t_o$, and $\sigma_i$ be the trailing annualized daily-return volatility. The anchor and terminal values are first averaged across modes and frameworks. All returns, shock magnitudes and volatility inputs below are numerical percentage points (12 denotes 12\%). The horizon-scaled threshold and dampening weight are:
\begin{equation*}
\begin{aligned}
s_i&=100\left|P(t)/P_{\mathrm{ref}}(t)-1\right|,\\
\theta_i&=\operatorname{clip}\!\left(\max(\sigma_i,5)\sqrt{d/365.25},8,20\right),\\
w_i&=\begin{cases}0,&s_i<12,\\0.5\max\!\left(\psi(s_i/\theta_i;0.75,1.5),\psi(s_i;12,25)\right),&s_i\ge12.\end{cases}
\end{aligned}
\end{equation*}
where $\psi(x;a,b)=3z^2-2z^3$ and $z=\mathrm{clip}((x-a)/(b-a),0,1)$. The guarded annualized return shrinks toward the 3.6\% cash hurdle:
\begin{equation*}
\widehat r^{\mathrm{guarded}}_{\mathrm{ann}}=3.6+(1-w_i)\left(\widehat r^{\mathrm{fixed}}_{\mathrm{ann}}-3.6\right).
\end{equation*}

Shape-Preserving Mechanical Reset ($\widehat{r}^{\mathrm{reset}}$): Keeps the old forecast's predicted relative growth rate $G_o(T)/G_o(t)$ from $t$ to $T$, but starts that growth path from the actual new price $P(t)$, shifting the implied terminal target to $P(t) G_o(T)/G_o(t)$. For example, if an old forecast started at 100, predicted 105 at $t$, and predicted 120 at $T$, but the actual price at $t$ is 110, fixed-target repricing yields $120/110 - 1 = 9.09\%$, whereas resetting yields $120/105 - 1 = 14.29\%$ (moving the implied target to 125.71; Figure~\ref{fig:baselines}). Because resetting discards the original valuation anchor, it is not a valid fundamental investment rule; however, it serves as an essential scientific control to test whether fresh LLM inference improves upon the slope of old forecasts or merely re-anchors their starting price level.

\subsection{Two Evaluation Lenses: Economic Portfolios vs. Statistical Accuracy}

To evaluate both real-world investment utility and statistical forecast mechanics, we test our update policies across two complementary setups:

\textbf{Lens 1 --- Economic Portfolio Decisions (Section 5):} Uses one-year annualized council forecasts across the 203 stocks to select equal-weight Top 10 (and Top 30) buy-and-hold portfolios in USD, comparing net returns against SPY over common 3-month and 6-month holding windows after FX conversion, net dividends, 10 bps per-leg trading costs, and upfront a USD 3,000 all-in research charge on a USD 300,000 account.

\textbf{Lens 2 --- Statistical Forecast Mechanics (Section 6):} Scores completed 90-day (first-quarter) targets across all 203 stocks in local quote currency using Mean Absolute Error (MAE), Root Mean Squared Error (RMSE), and rank overlap to isolate forecast accuracy and shortlist turnover from currency and dividend effects. For paired accuracy contrasts between fresh and reset forecasts, we compute:

\begin{equation}
\Delta = \frac{1}{N}\sum_{i,b}\left[\left\vert{}\widehat{r}^{\mathrm{fresh}}_{i,b} - r_{i,b}\right\vert{} - \left\vert{}\widehat{r}^{\mathrm{reset}}_{i,b} - r_{i,b}\right\vert{}\right], 
\end{equation}

where negative values favor fresh reports, accompanied by 95\% percentile intervals from 5,000 whole-stock bootstrap draws (resampling all 203 stock histories with replacement).

\section{Economic Evaluation: Can Price Updates Preserve Top 10 Investment Decisions?}

\subsection{One-Year Signal Construction and Portfolio Rules}

Our primary economic question is whether retaining a long-horizon forecast and updating its remaining upside can continue to support competitive Top 10 stock selections between deep refreshes. At each decision date---6 May, 8 June, and 6 July 2026---we aggregate the 10 matched council configurations across all 203 stocks. For Fixed, Guarded, and Fresh arms, implied one-year terminal prices are averaged across modes within each framework and then weighted equally ($1/7$) across the seven frameworks before annualizing the council return from the decision close. For Reset, per-report remaining growth factors are aggregated with the same framework weights before annualization. Under the native-endpoint specification, old forecasts have 323--339 days remaining to their original one-year target and fresh forecasts have 359--362 days remaining (we also test a strict 365-day common-endpoint interpolation in Section 5.3).

Within each arm, we rank the 203 stocks by annualized expected return (subject to a 3.6\% cash hurdle floor) and allocate equal starting capital across the Top 10 (and Top 30) slots. All 10 (and 30) slots fill in every window. All arms execute at the next common available market close (7 May--30 July, 9 June--4 September, and 7 July--30 September 2026, plus an extended 7 May--30 October 2026 six-month window for the first cohort) starting from a USD 300,000 account. Portfolio accounting uses the platform's production wallet engine, incorporating USD FX conversion, stock splits, dividend withholding, cash distributions, and accrued dividend receivables. An independent closed-form lot verifier reconciles terminal wealth across all 40 primary portfolio combinations to within USD $10^{-6}$.

The displayed baseline scenario deducts 10 basis points (bps) on entry and exit share turnover. The Fresh arm pays USD 3,000 upfront for a full 500-asset operating cycle, including retries, compute instances and infrastructure; the USD 300,000 account bears that full bill even though investment selections are evaluated only within the matched 203-stock panel. This is an all-in planning scenario, separate from the measured successful-report token reference in Section 7. Full results for 0, 10, and 25 bps with and without the research charge are provided in the data package. Normalized aggregate wealth paths independently reproduce all 240 original fee/cost cases before applying the revised capital and cost assumptions.

\begin{table}[!htbp]\centering\small
\caption{Top 10 buy-and-hold portfolio returns (USD), including all four signal arms and SPY, after 10 bps per-leg fees and a USD 3,000 upfront operating charge on Fresh and USD 300,000 starting capital.}\label{tab:portfolio}
\resizebox{\linewidth}{!}{\begin{tabular}{lrrrrr}
\toprule
Window & Old fixed & Old + guard & Path reset & Fresh & SPY \\
\midrule
3$\to$4 (~3mo) & 3.02\% & 2.44\% & 2.72\% & 4.55\% & 1.36\% \\
4$\to$5 (~3mo) & 4.45\% & 4.59\% & 22.92\% & 22.45\% & 4.47\% \\
5$\to$6 (~3mo) & 2.06\% & 2.06\% & 6.57\% & 3.93\% & 1.97\% \\
Same-model mean & 2.54\% & 2.25\% & 4.65\% & 4.24\% & 1.66\% \\
3$\to$4 (~6mo) & -6.15\% & -11.26\% & -3.98\% & 8.55\% & 3.46\% \\
\bottomrule
\end{tabular}
}
\end{table}

\begin{figure}[!htbp]\centering
\includegraphics[width=\linewidth]{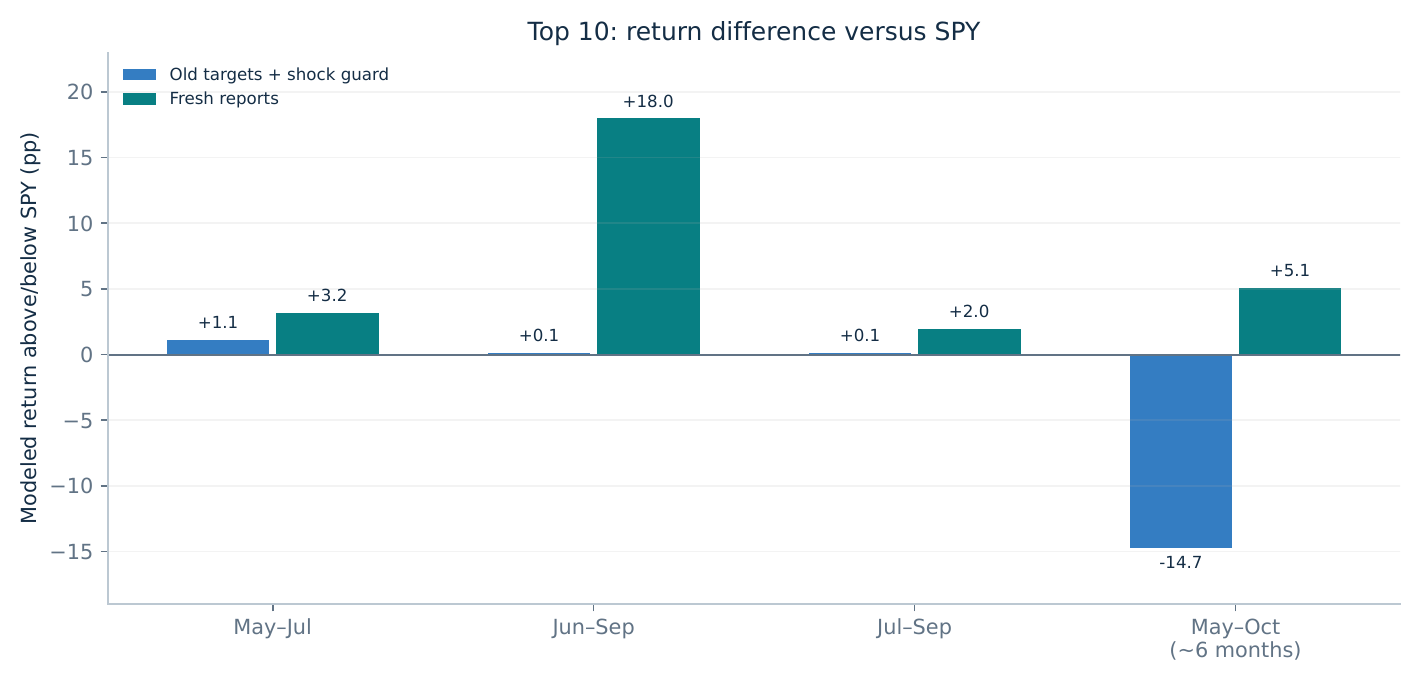}
\caption{Top 10 portfolio excess returns (in percentage points) relative to SPY over identical holding windows net of modeled costs. Guarded old targets outperform SPY across all three 3-month windows without new LLM inference, but suffer a sharp reversal when held unrefreshed for six months (May--Oct).}\label{fig:portfolio}
\end{figure}

\subsection{Short-Window Reuse and the Thesis Half-Life Question}

Table~\ref{tab:portfolio} and Figure~\ref{fig:portfolio} reveal substantial holding-horizon sensitivity:

\textbf{Three-Month Reuse Beats SPY at Zero Marginal Compute Cost:} Across all three 3-month holding windows, price-updated old targets with the production shock safeguard outperform SPY: 2.44\% vs. 1.36\% (May--Jul), 4.59\% vs. 4.47\% (Jun--Sep), and 2.06\% vs. 1.97\% (Jul--Sep). Across the two unchanged-model transitions ($3\to4$ and $5\to6$), guarded reuse averages 2.25\% (2.54\% unguarded) versus 1.66\% for SPY, while fresh reports average 4.24\% after the USD 3,000 operating charge on a USD 300,000 account (4.65\% for the mechanical path reset).

\textbf{Six-Month Neglect Breaks Down:} When the May cohort ($3\to4$) is held for nearly six months (176 days, through 30 October 2026) without an interim refresh or reranking, the guarded old-target Top 10 portfolio falls to -11.26\% (-6.15\% unguarded and -3.98\% reset), whereas the refreshed Top 10 portfolio returns +8.55\% versus +3.46\% for SPY. This longer-window reversal motivates the thesis half-life question: how long retained research continues to support useful investment decisions.

\textbf{Breadth Reversal (Top 30 vs. Top 10) and Drawdown Profile:} Higher concentrated returns in the Fresh Top 10 portfolio come with higher volatility, including a 16.8\% maximum peak-to-trough drawdown in the first window compared to 4.5\% for SPY and 9.3\% for Fresh Top 30. Expanding the shortlist to Top 30 also reverses the unchanged-model comparison: across $3\to4$ and $5\to6$, guarded reuse averages 4.17\% (5.70\% and 2.64\%), outperforming fresh reports at 1.37\% (5.39\% and -2.66\%; see Appendix D, Table~\ref{tab:top30}).

\textbf{Dividend Receivables and Accounting Verification:} In several international listings, ex-dividend entitlements accrued during the holding window lacked confirmed payment dates at liquidation (up to USD 958.34 per USD 300,000 account in guarded $3\to4$ Top 10). A cash-only sensitivity that strips unpaid receivables from terminal wealth preserves the Fresh-versus-Guarded ordering, but changes the narrow June--September Guarded Top 10 comparison with SPY: guarded reuse falls from 4.59\% to 4.39\%, below SPY at 4.47\%. As an implementation check, an independent closed-form lot calculation (verify\_portfolio\_lots.py) reconciles terminal wealth across all 40 primary portfolio configurations to within USD $10^{-6}$ (maximum discrepancy 7.28e-11 USD) of the production wallet engine.

\subsection{Stress Tests: Endpoint Alignment, Safeguard Symmetry, and Outlier Concentration}

Headline portfolio returns can mask fragility to horizon interpolation, risk-filter design, and individual outlier stocks. Keeping trade execution dates, liquidation dates, and fee rules fixed, we run four sensitivity checks (reported in full in Appendix D, Table~\ref{tab:robustness}):

\textbf{Common 365-Day Forecast Endpoint:} Under native endpoints, old reports have 323--339 days remaining to their original one-year target at the decision date, whereas fresh reports have 359--362 days. Interpolating every old and new path to a standardized decision date + 365 days horizon leaves Fresh Top 10 selections completely unchanged across all windows. However, it alters boundary selections in other arms: in July--September ($5\to6$), Reset Top 10 jumps from 6.57\% to 20.38\% after two stock swaps, and Fresh Top 30 rises from -2.66\% to -0.36\% after one swap. In June--September ($4\to5$), Guarded Top 10 drops from 4.59\% to -0.69\% when two positions change. Annualizing different native horizons does not fully eliminate term-structure effects.

\textbf{Symmetric Safeguard Application:} In our baseline setup, the D3 shock dampener is applied to old repriced targets while fresh reports are ranked on raw expected returns. Applying the exact same dampener symmetrically to fresh reports (measuring price-path deviation from the fresh forecast anchor to the decision close) triggers non-zero adjustments on 17 of 609 stock-date signals. Fresh Top 10 selections and returns remain 100\% identical in every window, while May--July ($3\to4$) Fresh Top 30 dips slightly from 5.39\% to 4.42\% (16.86\% to 15.91\% over six months).

\textbf{Volatility Input Stress Tests:} Because the historical price table used to compute 10-year trailing volatility lacks a split-adjusted close column, pre-2026 stock splits can inflate raw-close volatility estimates. To probe this sensitivity, we re-run the guarded policy under uniform annual volatility inputs of 5\% and 1000\%, forcing the shock threshold $\theta_i$ to its 8\% floor or 20\% ceiling across all stocks. In July--September ($5\to6$), Guarded Top 10 returns 5.00\% at the 5\% floor and -0.43\% at the 1000\% ceiling (vs. 2.06\% baseline), while June--September ($4\to5$) shifts to -0.44\% and 4.86\% (vs. 4.59\% baseline). Under both uniform stress scenarios, the two-window same-model average still favors Fresh Top 10 (4.24\%) over Guarded Top 10 (3.72\% at the floor and 1.30\% at the ceiling).

\begin{figure}[!htbp]\centering
\includegraphics[width=\linewidth]{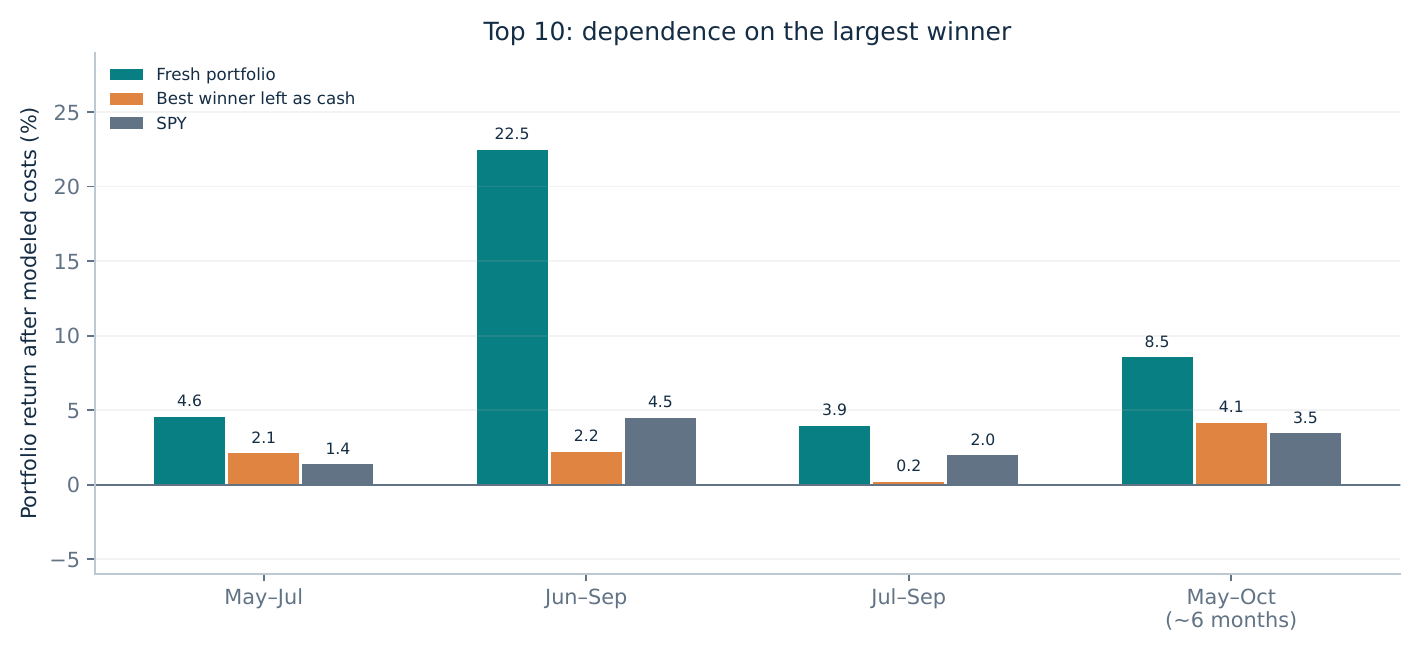}
\caption{Concentration stress test for Fresh Top 10 portfolios. Each bar group compares the baseline Fresh Top 10 return against the same portfolio with its single best-performing stock slot left in cash (retaining full upfront research charges) and against SPY.}\label{fig:concentration}
\end{figure}

\textbf{Single-Winner Concentration (MRNA Sensitivity):} To measure dependence on outlier winners, we remove each portfolio's single largest positive contributor ex-post, leaving that 1/k capital slot in zero-interest cash while retaining the full upfront compute fee (Figure~\ref{fig:concentration}). In the $4\to5$ transition, removing MRNA drops Fresh Top 10 from 22.45\% to 2.22\% (below SPY's 4.47\%) and Reset Top 10 from 22.92\% to 2.48\%. In the six-month $3\to4$ window, removing MRNA reduces Fresh Top 30 from 16.86\% to 5.12\%, below the original Guarded Top 30 return of 5.71\%. Across the two same-model short windows, removing each arm's own best winner lowers the Fresh Top 10 average from 4.24\% to 1.15\% (below SPY at 1.66\%) and Guarded Top 10 from 2.25\% to -0.28\%. The removed Fresh winners are PYPL in $3\to4$ and U in $5\to6$; the removed Guarded winners are RIVN and WDAY, respectively. Concentrated 10-stock portfolios across all update rules rely heavily on capturing right-tail winners.

\subsection{Improving the Ranking of Retained Research}
\label{sec:ranking-improvement}
The preceding comparisons evaluate the four declared update policies; their
results and research-cost question remain unchanged. We next ask whether a
better ranking rule can extract additional investment value from the same
retained forecasts. This is an incremental selection-policy experiment, not
new deep research or a replacement for the preceding evidence.

We retain the 203-stock panel, original decision and execution dates,
equal Top 10/30 slot allocations, dividend wallet, FX, corporate actions,
trading charges and all-in research bill. G10 ranks annualized expected
price return above the 3.6\% cash hurdle with bounded discounts for original
framework disagreement. Its continuity variant combines a fixed original
score $S_0$ with the current-price score $S_t$:
\[
 S_t^{75/25}=\max(0,0.75S_0+0.25S_t).
\]
The original score is evaluated from the original forecast inputs, not
yesterday's blended score. Current BUY economics still apply; the final
score must also pass the BUY threshold, with unfilled slots remaining
cash. We apply raw and blended G10 to \emph{both} reused and fresh report
councils. This controlled price-forecast test preserves the seven equally
weighted frameworks formed from ten reports and their frozen original
dispersion; it does not add forecast dividends or claim a complete replay
of the native fourteen-advisor V8 stack. Actual configured net dividends
and earned receivables remain included in realized wealth.

\begin{table}[!ht]
\centering\small
\begin{tabular}{lrrrrr}\toprule
Transition & Reuse raw & Reuse 75/25 & Fresh raw & Fresh 75/25 & SPY\\\midrule
3$\to$4 & 3.72 & 3.46 & 4.55 & 4.83 & 1.36\\
4$\to$5 & 7.42 & 27.16 & 25.86 & 26.97 & 4.47\\
5$\to$6 & 6.19 & 8.91 & 6.75 & 6.75 & 1.97\\
3$\to$4 (long) & -1.57 & -4.23 & 8.55 & 38.36 & 3.46\\
\bottomrule\end{tabular}
\caption{Matched Top 10 net holding-period returns (\%). USD300,000, 10 bps per actual share trade on entry and exit, USD3,000 incremental research bill on fresh arms, configured net realized dividends and earned receivables. The first three rows are short windows; the last holds the first selections longer. These returns are neither annualized nor compounded across rows.}
\end{table}

In the unchanged-model short windows ($3\to4$ and $5\to6$), raw G10 reuse
averages 4.96\% and blended reuse 6.19\%: an additional 1.23 percentage
points. Fresh raw and blended G10 average 5.65\% and 5.79\%, respectively,
after the USD3,000 research charge; SPY averages 1.66\%. The 1.23-point
contrast isolates the blend within G10. The larger increase from the
historical guarded arm's 2.25\% cannot all be attributed to continuity
because the base ranking family also changes. Replacing each policy's
best sleeve with cash at its original weight leaves 2.40\% for raw reuse
versus 3.07\% for blended reuse. Neither reuse policy holds Moderna in
these two short windows.

Against the same fresh-raw-G10 shortlist, reused Top 10 overlap rises
from 2/10, 3/10 and 2/10 to 3/10, 5/10 and 4/10 across the transitions.
When both sides use 75/25, overlaps are 2/10, 4/10 and 4/10. Full-panel
rank correlation slightly declines in the first two transitions and rises
in the third; closer shortlist agreement is neither universal rank
agreement nor proof of accuracy, since fresh research is a comparator.

The improvement has material boundaries. The $4\to5$ blended reuse return
of 27.16\% becomes 6.73\% with its Moderna sleeve held as cash, while
raw reuse did not hold Moderna. Across all three short windows that common
issuer control leaves a 0.59-point mean benefit. For Top 30, the blend
underperforms raw reuse in two of three short windows; the first blend
fills only 18 slots. In the longer Top 10 window, blended reuse returns
$-4.23$\% versus raw reuse's $-1.57$\%, although its drawdown is smaller
($-10.27$\% versus $-14.68$\%). Fresh raw G10 returns 8.55\% after
research cost. Fresh blended G10's 38.36\% drops to 3.13\% with Moderna
as cash. These controls retain the original allocation and research bill,
without adding a hindsight replacement stock.

Thus ranking continuity can improve concentrated reuse decisions without
eliminating the value of fresh research as a thesis ages. All 480 earlier
cost-version scenarios reproduce in the extension; 88 unit portfolios
generate 792 fee/cost scenarios, not 792 independent economic trials.
Full formulas, original-scoring versus rescoring experiments, risk-adjusted
alternatives and concentration checks are developed in the companion
manuscript \emph{Ranking AI Investment Forecasts: Original Scoring,
Rescoring, and Return-Focused versus Risk-Adjusted Strategies}.

\section{Statistical Forecast Mechanics: Why Cheap Baselines Compete}

\begin{figure}[!htbp]\centering
\includegraphics[width=\linewidth]{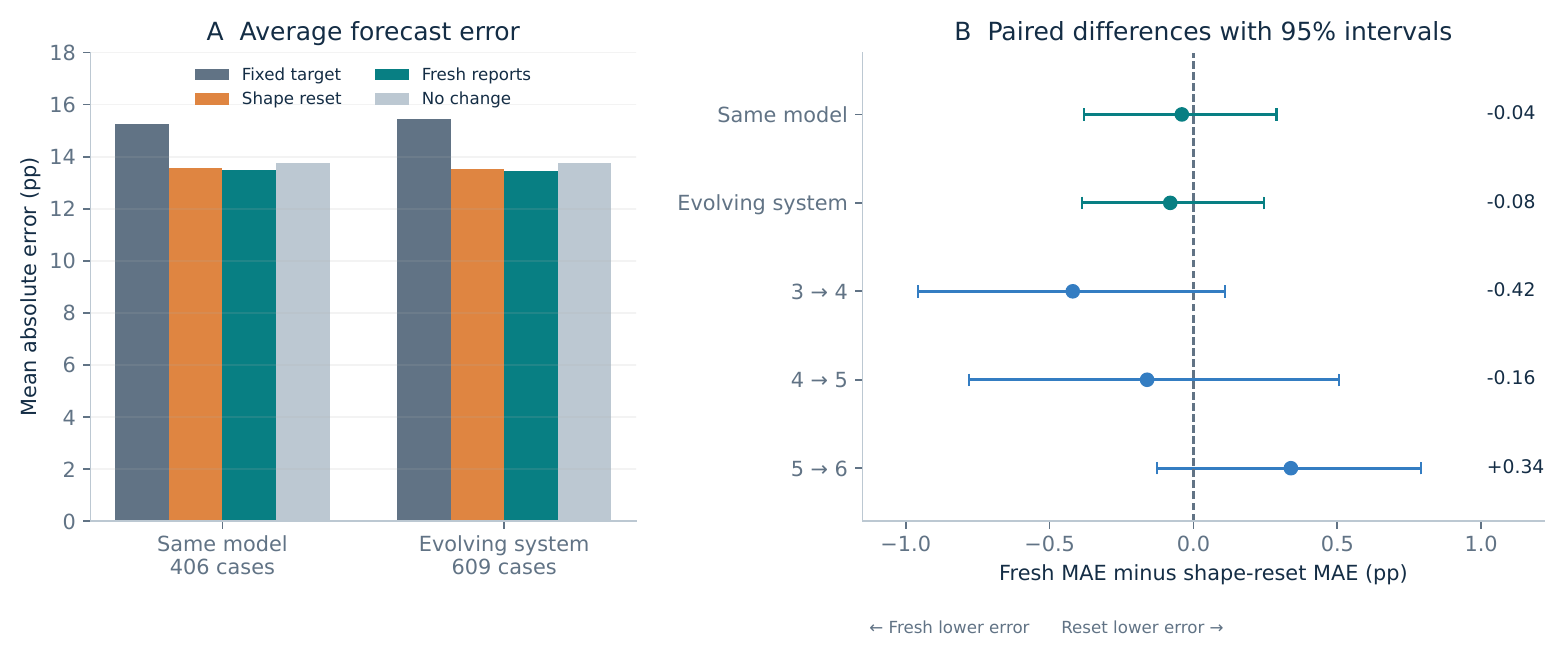}
\caption{First-quarter council forecast error across the 203-stock panel. Panel A: Mean Absolute Error (MAE) in percentage points across the same-model (406 cases) and evolving-system (609 cases) series. Panel B: Paired Fresh-minus-Reset MAE differences with 95\% whole-stock bootstrap percentile intervals. Negative values indicate lower error for fresh reports.}\label{fig:accuracy}
\end{figure}

To understand why zero-inference updates compete so closely with fresh LLM reports over three-month windows, we examine our second evaluation lens: completed first-quarter (85--89 day) price-return forecasts across all 203 stocks in local quote currency (Table~\ref{tab:primary} and Figure~\ref{fig:accuracy}).

\begin{table}[!htbp]\centering\small
\caption{Council first-quarter price-return forecast errors in percentage points. The same-model and evolving series contain 406 and 609 cases on the same 203 stocks.}\label{tab:primary}
\resizebox{\linewidth}{!}{\begin{tabular}{lrrrr}
\toprule
Method & Same-model MAE & RMSE & Evolving MAE & RMSE \\
\midrule
Fixed-target repricing & 15.257 & 22.708 & 15.431 & 23.552 \\
Shape-preserving reset & 13.548 & 19.428 & 13.519 & 20.489 \\
Fresh deep reports & 13.507 & 19.583 & 13.438 & 20.301 \\
No-change forecast & 13.746 & -- & 13.739 & -- \\
\bottomrule
\end{tabular}
}
\end{table}

\subsection{The Level-vs.-Slope Decomposition}

In the two same-model transitions, generating fresh deep reports reduces council MAE by 1.750 pp relative to fixed-target repricing (13.507 pp vs. 15.257 pp, 95\% stock-cluster bootstrap CI [-2.713, -0.831] pp). At first glance, that gap appears to validate paying for fresh LLM reasoning.

\textbf{However, comparing against the mechanical Shape-preserving reset tells a completely different story:} simply re-anchoring the old report's predicted growth rate to the new market price reduces MAE by 1.709 pp (13.548 pp, CI [-2.685, -0.768] pp), capturing 97.7\% of the MAE improvement without a single token of new inference. The remaining MAE difference between Fresh reports and Shape reset is just -0.040 pp (95\% CI [-0.380, +0.289] pp) in the same-model panel and -0.081 pp (95\% CI [-0.386, +0.244] pp) across the full 609-case evolving system. Under RMSE, the mechanical reset actually slightly outperforms fresh reports in the same-model panel (19.428 pp vs. 19.583 pp), and a naive 0\% no-change forecast achieves an MAE of 13.746 pp. Across individual transitions, the paired Fresh-minus-Reset MAE difference flips sign (-0.419 pp in $3\to4$, -0.161 pp in $4\to5$, and +0.339 pp in $5\to6$), with all three 95\% intervals crossing zero (Figure 5B).

\subsection{The Accuracy--Selection Paradox: Similar MAE, Massive Rank Churn}

Why can portfolio returns diverge sharply between methods when their average forecast errors across the 203 stocks are virtually identical? Because portfolio construction selects from the extreme upper tail of the cross-sectional distribution.

Even though Fresh and Reset differ by only 0.04 pp in pooled MAE, their mean absolute forecast revision is 2.762 pp (3.167 pp in the evolving series), producing cross-sectional Spearman rank correlations of 0.609, 0.591, and 0.733 across the three transitions. As a result, quarter-ranked Fresh and Reset Top 10 lists share only 1/10, 4/10, and 4/10 stocks (12/30, 15/30, and 9/30 for Top 30). When stocks are ranked purely by predicted next-quarter returns (Appendix B), Reset outperforms Fresh on Top 10 equal-weight price returns in both same-model windows (13.87\% vs. 4.98\% average), whereas Fresh outperforms Reset on Top 30 (6.00\% vs. 1.48\% average). Extending the $3\to4$ accuracy evaluation to the six-month (177--179 day) cutoff yields the same pattern: a small, statistically uncertain MAE difference (-0.360 pp, 95\% CI [-1.161, +0.457] pp, 20.135 pp vs. 20.495 pp) alongside opposite return rankings at Top 10 and Top 30 (Appendix B.3, Table~\ref{tab:longer}).

\section{Operating Budgets: Weekly, Monthly, or Reuse First?}

\subsection{Measured Tokens and the Full Operating Bill}

Across all four batches in the matched archive, the 8,120 distinct reports generated 46.11 million visible output tokens and 50.72 million reasoning tokens---96.83 million generated tokens in total---alongside 529.23 million input tokens. The token export counts each matched task once; failed calls, retries, synthesis and reports outside the matched panel are excluded.

Across Batches 4, 5, and 6, the 609 fresh stock councils (6,090 individual report runs) consumed 411.38 million input tokens and 72.13 million output-plus-thinking tokens. Table~\ref{tab:cost} retains the successful-report workload at a common API price schedule: USD 2.00 per million uncached input tokens, USD 0.20 cached input tokens and USD 12.00 output-plus-thinking tokens, with the batch tier halving uncached input and output rates \cite{pricing}. These normalized charges are a workload benchmark, not the full cost of operating the platform (Appendix E).

\begin{table}[!htbp]\centering\small
\caption{Measured successful-report tokens and normalized API reference charges for each 203-stock, ten-report refresh. These figures are retained separately from the all-in operating scenario.}\label{tab:cost}
\resizebox{\linewidth}{!}{\begin{tabular}{lrrrrr}
\toprule
New batch & Input M & Output M & Standard USD & Batch USD & Mean council USD \\
\midrule
4 & 147.31 & 25.25 & 473.87 & 243.81 & 2.33 \\
5 & 125.53 & 23.70 & 487.22 & 246.29 & 2.40 \\
6 & 138.55 & 23.18 & 525.17 & 264.26 & 2.59 \\
Total / mean & 411.38 & 72.13 & 1486.27 & 754.36 & 2.44 \\
\bottomrule
\end{tabular}
}
\end{table}

Our main operating scenario budgets \textbf{USD 3,000 per 500-asset run}, including model inference, retries, compute instances, infrastructure and other operating overhead. This owner-specified planning assumption replaces the token-only bill in the primary economic comparison. On a \textbf{USD 300,000 portfolio}, one full run consumes 1\% of starting capital. Shared-service allocation provides a secondary comparison: allocating the same cycle proportionally to 203 of 500 assets costs USD 1,218. We charge the full USD 3,000 in the main test without assigning any selection benefit to the other 297 assets.

\subsection{The Capital-Scaled Break-Even Cost}

Let $C$ denote starting capital, $R$ the incremental upfront research bill, and $W_f$ and $W_o$ the fresh and guarded-reuse terminal wealth factors after trading fees but before research costs. Then:
\begin{equation}
V_f=(C-R)W_f,\qquad V_o=CW_o,
\end{equation}
\begin{equation}
R_{\max}^{\mathrm{realized}}=C\left(1-\frac{W_o}{W_f}\right),\qquad
C_{\min}^{\mathrm{break\text{-}even}}=\frac{R}{1-W_o/W_f}\quad(W_f>W_o).
\end{equation}
These quantities connect a fixed bill to the capital pool it serves. For the USD 300,000 Top 10 account, the realized research ceilings in the two unchanged-model short windows are USD 8,990 (May--Jul) and USD 8,326 (Jul--Sep). A single USD 3,000 run fits below both ceilings. The minimum capital supporting that one-off bill is approximately USD 100,115 and USD 108,093, respectively. At Top 30, the May--Jul ceiling contracts to USD 2,126 and the Jul--Sep ceiling is negative: broader fresh selections do not earn back the full operating bill in either unchanged-model window.

\begin{table}[!htbp]\centering\small
\caption{Capital-sharing sensitivity for one USD 3,000 full-cycle charge. Trading fees remain 10 bps per leg; the same selections and realized market outcomes are retained.}\label{tab:budget}
\resizebox{\linewidth}{!}{\begin{tabular}{lrrrr}
\toprule
Transition & Capital USD & Research USD & Guarded reuse & Fresh \\
\midrule
3$\to$4 & 10,000 & 3000.00 & 2.44\% & -26.07\% \\
3$\to$4 & 100,000 & 3000.00 & 2.44\% & 2.44\% \\
3$\to$4 & 300,000 & 3000.00 & 2.44\% & 4.55\% \\
3$\to$4 & 1,000,000 & 3000.00 & 2.44\% & 5.29\% \\
5$\to$6 & 10,000 & 3000.00 & 2.06\% & -26.52\% \\
5$\to$6 & 100,000 & 3000.00 & 2.06\% & 1.83\% \\
5$\to$6 & 300,000 & 3000.00 & 2.06\% & 3.93\% \\
5$\to$6 & 1,000,000 & 3000.00 & 2.06\% & 4.66\% \\
\bottomrule
\end{tabular}
}
\end{table}

\subsection{Weekly Versus Monthly: What Must the Extra Runs Earn?}

A single refresh and a recurring refresh program impose very different hurdles. At USD 3,000 per run, 52 weekly cycles cost USD 156,000---52\% of the USD 300,000 starting portfolio. Twelve monthly cycles cost USD 36,000, or 12\%. Sharing research across a larger capital pool reduces both fractions proportionally.

For each observed window, Table~\ref{tab:cadence} counts an initial run at entry, then every seven calendar days or on the same date each month, excluding a run at liquidation. The short windows require 12--13 weekly runs or three monthly runs; the 176-day extension requires 26 weekly or six monthly runs. To isolate the budget hurdle, we reserve the entire scheduled bill upfront in zero-interest cash and retain the initial Fresh Top 10 holdings. Thus $V_f^{\mathrm{budget}}=(C-nR)W_f$, where $n$ is the number of scheduled cycles. The archive contains no weekly fresh reports, so this calculation measures the value additional runs would need to create; it does not invent their forecasts or interim trades.

\begin{table}[!htbp]\centering\small
\caption{Weekly and monthly research-budget sensitivity on USD 300,000. ``Fresh held'' retains the initial fresh-selected stocks after reserving the full scheduled budget. Reuse pays no new inference charge. These are fixed-selection cost scenarios, not observed recurring-refresh strategies.}\label{tab:cadence}
\resizebox{\linewidth}{!}{\begin{tabular}{lrrrrrr}
\toprule
Window & Budget cadence & Runs & Budget USD & Capital & Fresh held & Reuse \\
\midrule
05-07--07-30 & Weekly & 12 & 36,000 & 12\% & -7.06\% & +2.44\% \\
05-07--07-30 & Monthly & 3 & 9,000 & 3\% & +2.44\% & +2.44\% \\
06-09--09-04 & Weekly & 13 & 39,000 & 13\% & +7.61\% & +4.59\% \\
06-09--09-04 & Monthly & 3 & 9,000 & 3\% & +19.98\% & +4.59\% \\
07-07--09-30 & Weekly & 13 & 39,000 & 13\% & -8.67\% & +2.06\% \\
07-07--09-30 & Monthly & 3 & 9,000 & 3\% & +1.83\% & +2.06\% \\
05-07--10-30 & Weekly & 26 & 78,000 & 26\% & -18.86\% & -11.26\% \\
05-07--10-30 & Monthly & 6 & 18,000 & 6\% & +3.06\% & -11.26\% \\
\bottomrule
\end{tabular}
}
\end{table}

\begin{figure}[!htbp]\centering
\includegraphics[width=\linewidth]{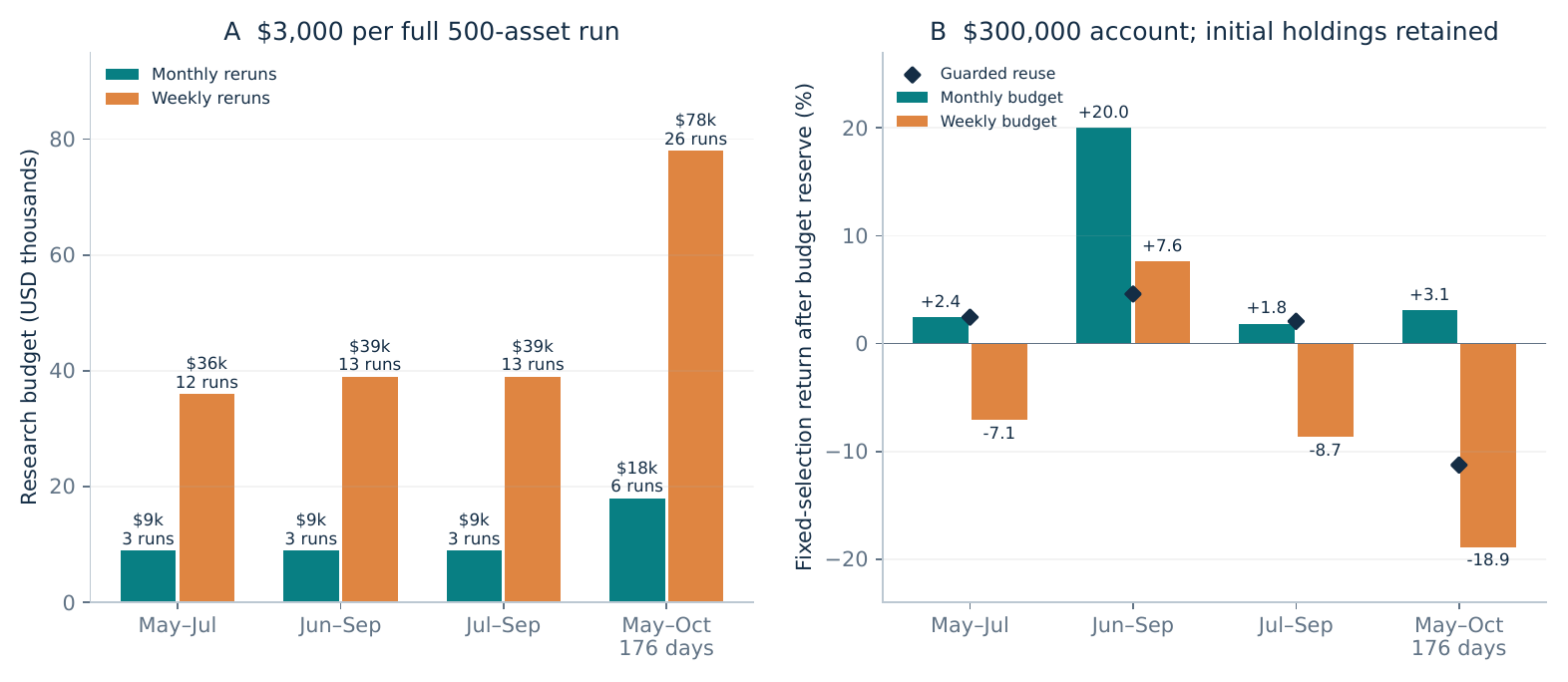}
\caption{The cost hurdle of recurring research. Panel A shows weekly and monthly full-run budgets across the actual holding windows. Panel B deducts those reserved budgets while holding the initial Fresh Top 10 selection fixed; diamonds show guarded reuse. Additional reruns must improve decisions enough to overcome this hurdle.}\label{fig:cost}
\end{figure}

In the two unchanged-model short windows, monthly budgets leave fixed-selection returns of 2.44\% and 1.83\%, compared with 2.44\% and 2.06\% for guarded reuse. The May difference is less than 0.01 percentage points; in July the extra runs would need to create about USD 707 of terminal wealth to match reuse. Weekly budgets reduce the same initial selections to -7.06\% and -8.67\%, leaving gaps of approximately USD 28,526 and USD 32,200 to close. The evolving-system June window is stronger, but remains sensitive to MRNA's contribution (Section 5.3).

The longer window changes the decision. Over 176 days, one fresh run returns 8.55\% and guarded reuse loses 11.26\%. Reserving six monthly runs leaves the initial Fresh selection at 3.06\%; reserving 26 weekly runs reduces it to -18.86\%. This contrast identifies the operational objective: preserve useful research between genuine evidence changes, rather than pay for maximum refresh frequency. The available windows support that budget question without identifying a universal thesis half-life or an optimal schedule.

\section{Auditing Editorial Screening and Published-Pick Turnover}

\subsection{Publication-Bound Audit: Changing Editions and Confirmed Input Lists}

In live production, iPulse AI passes its quantitative shortlist through an editorial workflow designed to check dated news and flag any stock whose thesis has been invalidated (NEEDS\_REANALYSIS). We audit the platform's production decision records from August to October 2026. Excluding six failed publication attempts and one legacy shadow-mode run, the ledger contains nine governed stock-buy editions across Batches 6 and 7, with 90 recorded candidate positions in total.

\begin{table}[!htbp]\centering\small
\caption{New stocks relative to the preceding published edition versus replacements by that edition's final screen. Ten candidates were recorded in each edition.}\label{tab:editorial}
\resizebox{\linewidth}{!}{\begin{tabular}{lrrr}
\toprule
Edition (UTC date) & Batch & New vs prior & Screen replaced \\
\midrule
2026-08-27 & 6 & -- & 0 \\
2026-09-07 & 6 & 2/10 & 0 \\
2026-09-21 & 7 & 8/10 & 0 \\
2026-10-01 & 7 & 4/10 & 0 \\
2026-10-05 & 7 & 4/10 & 0 \\
2026-10-12 & 7 & 2/10 & 0 \\
2026-10-16 & 7 & 6/10 & 0 \\
2026-10-24 & 7 & 5/10 & 0 \\
2026-10-31 & 7 & 4/10 & 0 \\
\bottomrule
\end{tabular}
}
\end{table}

Across all nine editions, the post-review published Top 10 matches that edition's incoming quantitative Top 10 in exact rank order, with zero recorded NEEDS\_REANALYSIS flags or candidate replacements. Published membership nevertheless changes in every consecutive pair of editions: two to eight stocks enter each new list. The 21 September edition retains only 2/10 stocks from 7 September, while still matching its own contemporaneous Batch 7 input list 10/10.

\begin{figure}[!htbp]\centering
\includegraphics[width=\linewidth]{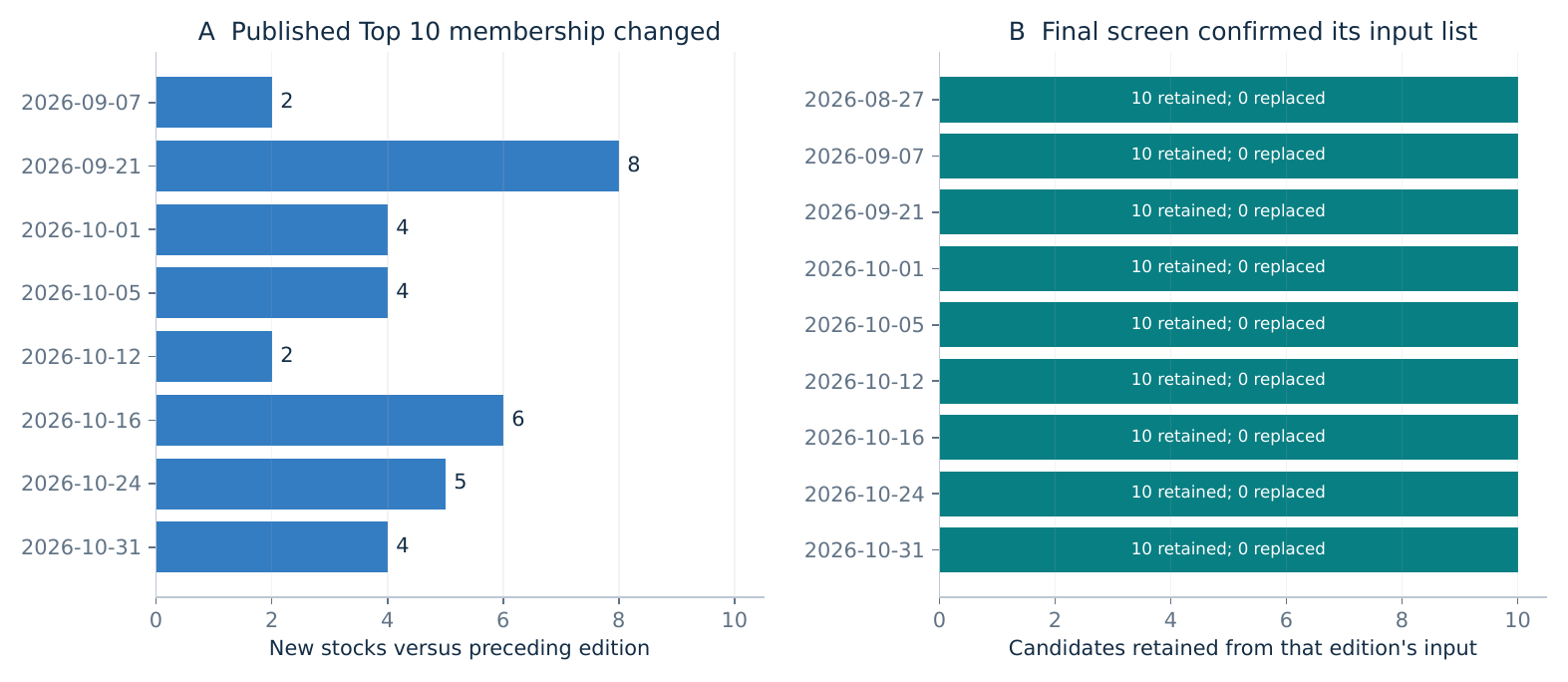}
\caption{Published-pick turnover versus final editorial validation. Panel A shows new stocks in successive published editions; Panel B shows that each final screen retains its own incoming ten candidates. These are distinct comparisons.}\label{fig:editorial}
\end{figure}

\textbf{These are two different comparisons:} turnover between published editions and replacements by the final screen within one edition. Recording both exposes where membership changes enter the workflow and provides a direct basis for evaluating future editorial interventions.

\section{A Budget-Constrained Architecture for Agentic Research}

Section~\ref{sec:ranking-improvement} adds a practical refinement: better ranking can improve the return extracted from retained forecasts before paying for another research run. That improvement changes a reuse policy, not the underlying research-budget question. Its breadth, concentration and longer-window limits preserve the need to assess whether new fundamental evidence can justify the next deep refresh.

\subsection{Formalizing the Three-Tier Allocation Problem}

Consider a research platform covering a universe of $M$ equities under a per-period compute and operational budget $B_t$. At each decision date $t$, the platform updates market prices across all $M$ assets at cost $C_{\mathrm{price},t}$, selects a subset $D_t \subseteq \{1,\dots,M\}$ for deep multi-agent refreshes at unit cost $c_{i,t}^D$ (including all agent runs and synthesis), and screens a candidate shortlist $E_t$ editorially at unit cost $c_{i,t}^E$, subject to:

\begin{equation}
C_{\mathrm{price},t} + \sum_{i \in D_t} c_{i,t}^D + \sum_{i \in E_t} c_{i,t}^E + C_{\mathrm{shared},t} \le B_t.
\end{equation}

Under a decision-focused objective \cite{spo}, refreshing subset $D_t$ is economically justified only when the expected incremental portfolio utility of replacing stale signals in $D_t$ exceeds the incremental compute expenditure $\sum_{i \in D_t} c_{i,t}^D$ conditional on information available at $t$.

A platform can optimize three different outcomes with that budget: forecast accuracy across the full universe, the quality of the shortlisted stocks, or the net utility of the portfolio actually held. These objectives place value on different information. Revising a forecast for a stock far below the selection boundary may improve aggregate MAE without changing an investment decision. A smaller revision near the Top 10 cutoff can change the portfolio even when its effect on universe-wide MAE is negligible. The two evaluation lenses in this paper make that distinction visible.

The budget should follow the service being delivered. For an investor sharing a platform-wide research bill, the relevant charge is the allocated share of report generation, synthesis, retrieval and screening. For a stand-alone account funding a full refresh, the same bill falls on a much smaller capital base. Section 7 quantifies the latter comparison; the allocation rule above makes the shared costs explicit. Editorial evidence acquisition and review time belong in their own cost ledger rather than being assigned an arbitrary fraction of a deep council's bill.

Coverage and frequency introduce a separate choice. Analyzing 200 stocks four times and 400 stocks twice both produce 800 stock-analysis episodes, but allocate attention differently. Broader coverage may uncover a better candidate; more frequent work may update a thesis already held. Actual costs also depend on model choice, report length, caching and synthesis. A budget-matched breadth-versus-frequency experiment would complement the fixed-universe comparison here.

\subsection{Practical Operating Prescription: Reuse-First, Trigger-on-Evidence}

Our empirical findings point directly to a reuse-first, trigger-on-evidence operating architecture for multi-agent financial systems:

\textbf{Treat Deep Reports as Durable Assets with Daily Guarded Repricing:} Maintain continuous daily price updates across the full universe using the D3 shock safeguard to prevent broken theses from rising to the top of value rankings between deep runs.

\textbf{Test Periodic Refresh Schedules:} The short-window results and unfavorable longer extension motivate comparing scheduled full-universe refreshes under a fixed budget. Quarterly full-universe refreshes offer a candidate cadence for that prospective comparison.

\textbf{Replace Blanket Re-Runs with Event-Triggered Nominators:} Between scheduled batches, use large price-path shocks ($w_i > 0$), earnings releases, and material news flags to nominate evidence reviews and determine when a full-batch refresh is justified---and log every editorial rejection, replacement, and no-change decision so that downstream screening rules can be audited against pre-review baselines.

\textbf{Make the Editorial Layer an Evidence Check:} Preserve the incoming quantitative order, record a decision for each candidate reviewed, and retain the evidence behind both confirmations and invalidations. If a leading candidate's thesis is invalidated, consider the next eligible reserve from the frozen ranking. This gives the screening layer a clear purpose: update the shortlist when material evidence changes its investment case, while avoiding another full-universe reasoning run for an unchanged thesis.

For a prospective comparison, record the action as well as its result: the cost of retaining a report, reviewing a candidate, or rebuilding a council; the shortlist before and after the action; and the portfolio subsequently executed. Those records connect resource allocation to investor outcomes and make an unchanged shortlist informative alongside a replacement.

\subsection{Protocol for a Prospective Confirmatory Trial}

To move from retrospective archival diagnostics to a confirmatory test of optimal refresh scheduling, future evaluations should pre-register six elements prior to market realizations:

\textbf{Frozen Universe and Actionable Clocks:} Declare the eligible asset universe and bind all trade signals to public ledger availability timestamps (publication\_available\_at\_utc), retaining delisted or split stocks via point-in-time corporate-action adjustments.

\textbf{Saved Starting Artifacts:} Archive exact prompt packages, anchor prices, forecast curves, and model release IDs while keeping fixed-target repricing, guarded repricing, and mechanical path resets distinct.

\textbf{Budget-Matched Treatment Arms:} Compare scheduled full-universe refreshes against event-triggered partial refreshes ($D_t$) and editorial screening ($E_t$) under an identical dollar compute cap $B_t$.

\textbf{Predefined Non-Inferiority and Risk Endpoints:} Specify primary portfolio size (e.g., Top 10), maximum drawdown limits, transaction fees, and an acceptable return shortfall margin $\delta$ relative to continuous refreshing before observing returns.

\textbf{Auditable Editorial Logs:} Record pre-review candidate rankings, dated news evidence, rejection reasons, replacement IDs, and review token costs for every edition.

\textbf{Temporal Power:} Accumulate sufficient independent refresh dates across market regimes to estimate time-series standard errors alongside cross-sectional stock bootstrap intervals.

\section{Scope Conditions and Methodological Boundaries}

To keep the empirical sections focused, we consolidate the boundaries of our archival design here:

\textbf{Retrospective Design and Sample Horizon:} While all 8,120 reports were generated in live production prior to market realizations (with archived generation timestamps preceding the evaluated outcomes), the experimental comparisons, Top 30 sensitivity checks, and stress tests were designed retrospectively. Furthermore, while the panel spans 203 stocks and 6,090 report pairs, those observations occur across three overlapping refresh dates (May, June, and July 2026) evaluated through October 2026. Whole-stock bootstrap intervals capture cross-sectional sampling variation conditional on these market windows, not generalizability across future macroeconomic regimes or multi-year holding horizons.

\textbf{System Evolution vs. Pure Model Isolation:} Batches 3--4 (20251119\_preview\_release) and Batches 5--6 (20260201\_preview\_release) preserve fixed model release identifiers within each pair, whereas the $4\to5$ transition reflects a combined platform upgrade (model release, equity fundamentals injection, and schema v9). Even within same-model pairs, live market context and prompt hardening evolved; thus, transitions measure realistic production refreshes rather than sterile single-variable ablations.

\textbf{Simplified Portfolio Construction vs. Full Production Stack:} To isolate the marginal effect of refreshing council forecasts, our portfolio backtest ranks stocks solely by equal-framework council returns (plus the D3 shock guard) under simple equal-slot buy-and-hold rules. It does not replay the live platform's multi-factor risk scoring, cross-horizon filters, or dynamic synthesizer weights, and uses saved daily closes without modeling intraday market impact or slippage.

\textbf{Operating Budget and Cadence Scenarios:} USD 3,000 per 500-asset run is an owner-specified all-in planning assumption, not a reconstructed historical invoice or measured component ledger. The main USD 300,000 account bears that full bill while selections remain confined to 203 matched stocks. The token benchmark excludes non-report costs. Weekly/monthly scenarios reserve the full scheduled bill upfront and retain initial holdings; no intermediate reports, trades, or refresh benefits are observed. Linearity assumes no market impact, fixed subscriptions or capital-dependent constraints.

\textbf{Cohort, Volatility, and Thesis Half-Life:} The common 203-stock panel still depends on retrospective coverage and split exclusions. The trailing volatility inputs use raw closes; full historical split normalization is unverified. Uniform floor/ceiling stress scenarios do not bound every stock-specific correction. The extended holding window motivates the thesis half-life concept but does not estimate a universal decay constant or an optimal quarterly refresh schedule.

\textbf{Editorial Identification:} The audit measures recorded input/output selections rather than the causal value of editorial research. Some historical helper submissions automatically populated VALID decisions; the zero-replacement record is therefore not an independent test of screening quality. It also does not causally allocate upstream turnover among price, batch and eligibility changes.

\section{Conclusion}

When operating a multi-agent financial research platform, rerunning deep LLM inference across the entire stock universe every time prices move is an expensive default---not an automatic improvement. Across 8,120 archived production reports on 203 stocks, we find that long-horizon AI research retains substantial short-term decision value when paired with disciplined price updating:

Over three-month windows, repricing existing one-year targets with a volatility shock safeguard beats SPY in every window (2.25\% vs. 1.66\% in same-model runs) at zero marginal LLM compute cost.

On 90-day forecast accuracy, 98\% of the MAE improvement from fresh LLM reports (1.75 pp over fixed targets) is replicated for free by mechanically resetting old growth paths to current prices (0.04 pp gap between fresh and reset).

\textbf{The thesis half-life question remains:} leaving the May Top 10 portfolio unrefreshed for six months results in a -11.26\% loss versus +8.55\% for refreshed research. Moreover, tiny differences in average forecast error mask 60\%--90\% turnover in Top 10 rankings, making realized portfolio gains highly sensitive to shortlist size (Top 10 vs. Top 30), capital scale, and single-stock outliers (MRNA).

A USD 3,000 full-universe run costs 1\% of a USD 300,000 account. Recurring weekly or monthly runs must create value beyond the initial research to earn their recurring bill. The fixed-selection budget sensitivities make that hurdle explicit across both short and longer windows.

\textbf{For builders of agentic research systems, the prescription is clear:} decouple continuous price monitoring from episodic thesis rebuilding. Treat deep multi-agent reports as durable assets whose valuation upside is updated daily with price-shock safeguards, log downstream editorial interventions against frozen pre-review baselines, and reserve costly deep refreshes for scheduled horizon expirations or event-driven thesis breaks.

\section*{Data, Code, and Reproducibility}

The complete paper and public research artifacts for Version 2.1.0 are archived at \href{https://doi.org/10.5281/zenodo.23941195}{doi:10.5281/zenodo.23941195}. Versioned code, forecast-only datasets and reproduction instructions are available in the \href{https://github.com/TheFutureEdge/research-when-fresh-ai-stock-analysis-is-worth-it}{public research repository}.

The accompanying research-artifacts package contains all aggregate result JSONs, the 203-stock cohort definition, sanitized portfolio and editorial selection records, figure and table generation scripts, and standalone verification adapters (recompute\_public.py, recompute\_portfolio\_signals.py, recompute\_selections.py, and verify\_portfolio\_lots.py). While raw licensed market price series, FX rates, and the full production wallet engine remain private due to data-redistribution restrictions, the public adapters allow independent verification of all signal aggregations, editorial identity checks, and terminal-lot wealth calculations when paired with separately licensed market data under the documented JSON schemas. The version-2 aggregate wealth paths also reproduce all 240 original fee/cost cases before repricing the same selections for the revised operating bill, account size and cadence budgets. Original prose and public research exports are released under CC BY 4.0, and the reproduction scripts under the MIT License.

Version 2.1.0 adds the controlled ranking-policy comparison in Section~\ref{sec:ranking-improvement}. Its derived signal inputs, aggregate wealth paths, outcomes and terminal concentration controls are in \texttt{data/ranking\_extension/}; \texttt{code/verify\_ranking\_extension.py} reproduces the published arithmetic. Licensed raw market observations and private wallet ledgers remain excluded.

\section*{Author Responsibility and Competing Interests}

Russlan Ramdowar is the founder of Future Edge Group FZE, which develops and operates the iPulse AI platform evaluated in this study. Historical equity forecasts in the archive were generated by production Gemini configurations (Gemini 3 Pro Preview and Gemini 3.1 Preview), and AI coding/research tools (including OpenAI Codex) assisted with audit scripts, data verification, and manuscript preparation. Full responsibility for the study design, analysis, and final text rests with the human author. This paper is an empirical systems evaluation and does not constitute investment advice.

\clearpage\appendix
\section{Frameworks, Modes, and Council Aggregation}

The ten fixed agent configurations spanning seven investment frameworks are listed below. All archived runs use standard temperature/creativity and a high reasoning-effort (thinking) budget, where Researcher and Thinker represent distinct execution modes within a given framework.

\begin{center}\small
\begin{tabular}{p{.25\linewidth}p{.20\linewidth}p{.44\linewidth}}\toprule
Framework label & Matched modes & Prompted perspective (brief)\\\midrule
Warren Buffett & Researcher, thinker & Business quality, valuation, and durable cash generation.\\
Ray Dalio & Researcher, thinker & Macro conditions, cycles, and cross-market relationships.\\
Superintelligence & Researcher, thinker & Broad synthesis across available evidence.\\
Niccolo Machiavelli & Researcher & Power, incentives, and institutional behavior.\\
Elon Musk & Researcher & Technological change and long-horizon disruption.\\
Sherlock Holmes & Researcher & Evidence checks, inconsistencies, and alternative explanations.\\
Michael Burry & Researcher & Contrarian valuation and downside vulnerabilities.\\\bottomrule
\end{tabular}
\end{center}

For forecast method $m$, let $\mathcal{F}$ denote the seven frameworks ($|\mathcal{F}| = 7$) and $\mathcal{M}_f$ the matched modes within framework $f$ ($|\mathcal{M}_f| \in \{1, 2\}$). The standardized council forecast for stock $i$ in transition $b$ is:

\begin{equation}
\widehat{r}_{i,b}^{m} = \frac{1}{7}\sum_{f \in \mathcal{F}} \frac{1}{|\mathcal{M}_f|}\sum_{k \in \mathcal{M}_f} \widehat{r}_{i,b,f,k}^{m}.
\end{equation}

Because non-linear ratios and arithmetic averages do not commute, each report's mechanical reset ratio $G_o(T)/G_o(t)$ is computed at the individual report level before averaging across modes and frameworks.

\section{Detailed Forecast Error and Shortlist Diagnostics}

\subsection{Transition-Level MAE and Win Shares}

While fresh reports beat fixed-target repricing by 1.750 pp in the same-model series (1.993 pp in the evolving series), the Shape-preserving reset closes nearly the entire gap (1.709 pp and 1.912 pp, respectively). Table~\ref{tab:paired} breaks down the paired Fresh-minus-Reset MAE contrast across pooled series and individual transitions.

\begin{table}[!htbp]\centering\small
\caption{Paired Fresh-minus-Reset first-quarter MAE differences, stock-cluster intervals and fresh win shares.}\label{tab:paired}
\resizebox{\linewidth}{!}{\begin{tabular}{lrrr}
\toprule
Series (cases) & $\Delta$ MAE (pp) & 95\% stock-cluster interval & Fresh wins \\
\midrule
Same model (406) & -0.040 & [-0.380, +0.289] & 49.5\% \\
Evolving (609) & -0.081 & [-0.386, +0.244] & 51.9\% \\
3$\to$4 (203) & -0.419 & [-0.958, +0.109] & 52.2\% \\
4$\to$5 (203) & -0.161 & [-0.779, +0.504] & 56.7\% \\
5$\to$6 (203) & +0.339 & [-0.126, +0.790] & 46.8\% \\
\bottomrule
\end{tabular}
}
\end{table}

The transition-level results explain why the pooled difference is so small: Fresh improves on Reset by 0.419 pp in $3\to4$, while Reset improves on Fresh by 0.339 pp in $5\to6$. Fresh wins 49.5\% of the same-model stock-transition cases and 51.9\% across the evolving series. Pooling therefore combines changes in opposite directions rather than a uniform accuracy improvement across stocks.

The benchmark also changes the apparent scale of the benefit. A no-change forecast has MAE 13.746 pp in the same-model series and 13.739 pp in the evolving series. Fresh beats the old fixed targets much more clearly than it beats this simple reference. Under same-model RMSE, Reset scores 19.428 pp against Fresh at 19.583 pp: typical absolute error and large-error sensitivity can favor different update rules.

\subsection{Cross-Sectional Rank Stability and Quarter-Ranked Shortlist Returns}

\begin{figure}[!htbp]\centering
\includegraphics[width=\linewidth]{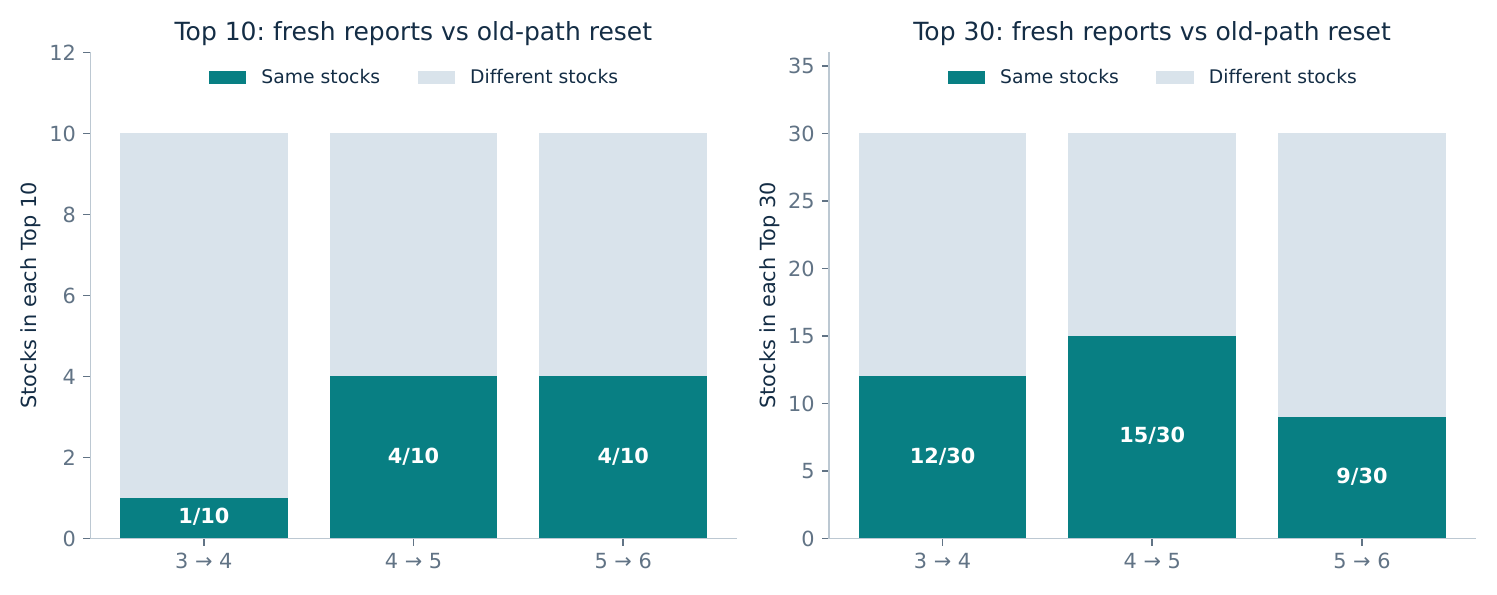}
\caption{Shortlist overlap between Fresh reports and Old-path reset for Top 10 (left) and Top 30 (right) selections ranked by predicted first-quarter return.}\label{fig:choices}
\end{figure}

Table~\ref{tab:rank} and Figure~\ref{fig:choices} show that despite near-zero pooled MAE differences, Fresh and Reset rankings diverge substantially in the upper tail.
\begin{table}[!htbp]\centering\small
\caption{Cross-sectional rank correlation and shared membership between Fresh and Reset quarter-ranked shortlists.}\label{tab:rank}
\resizebox{\linewidth}{!}{\begin{tabular}{lrrr}
\toprule
Transition & Rank correlation & Shared Top 10 & Shared Top 30 \\
\midrule
3$\to$4 & 0.609 & 1/10 & 12/30 \\
4$\to$5 & 0.591 & 4/10 & 15/30 \\
5$\to$6 & 0.733 & 4/10 & 9/30 \\
\bottomrule
\end{tabular}
}
\end{table}

\begin{figure}[!htbp]\centering
\includegraphics[width=\linewidth]{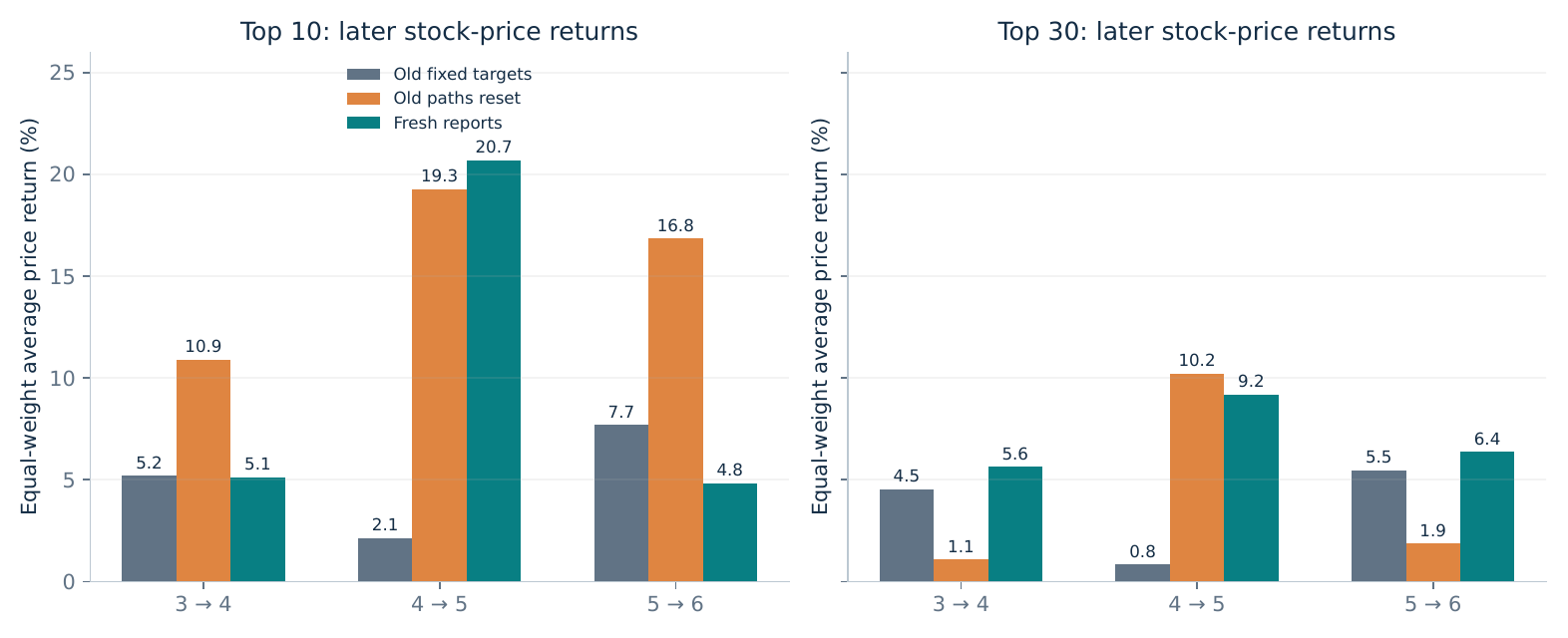}
\caption{Subsequent equal-weight native-quarter price returns for Top 10 (left) and Top 30 (right) quarter-ranked shortlists.}\label{fig:selection}
\end{figure}

Table~\ref{tab:selection_returns} and Figure~\ref{fig:selection} evaluate the realized equal-weight native-quarter price returns (in local quote currency, before dividends, FX, or fees) of the stocks selected by these first-quarter rankings.
\begin{table}[!htbp]\centering\small
\caption{Subsequent equal-weight native-quarter quote-currency price returns, before dividends, FX and fees. Column order is Old targets, Reset, Fresh.}\label{tab:selection_returns}
\resizebox{\linewidth}{!}{\begin{tabular}{lrrrr}
\toprule
Transition & Stocks chosen & Old targets & Old paths reset & Fresh reports \\
\midrule
3$\to$4 & 10 & 5.20\% & 10.89\% & 5.13\% \\
3$\to$4 & 30 & 4.55\% & 1.09\% & 5.63\% \\
4$\to$5 & 10 & 2.14\% & 19.27\% & 20.68\% \\
4$\to$5 & 30 & 0.85\% & 10.22\% & 9.19\% \\
5$\to$6 & 10 & 7.69\% & 16.85\% & 4.83\% \\
5$\to$6 & 30 & 5.48\% & 1.87\% & 6.37\% \\
Same-model mean & 10 & 6.45\% & 13.87\% & 4.98\% \\
Same-model mean & 30 & 5.01\% & 1.48\% & 6.00\% \\
\bottomrule
\end{tabular}
}
\end{table}

Tracing the exact stock replacements explains the divergence between Top 10 and Top 30 in the two same-model transitions:

In $3\to4$ Top 10, the 9 stocks added by Fresh returned 3.38\% on average, whereas the 9 stocks dropped from Reset returned 9.77\%. In $5\to6$ Top 10, the 6 Fresh additions returned 1.06\% versus 21.08\% for the 6 Reset removals.

Conversely, at Top 30, Fresh additions outperformed Reset removals in both same-model windows (4.51\% vs. -3.06\% in $3\to4$; 5.03\% vs. -1.40\% in $5\to6$).

Averaging the two same-model windows, the quarter-ranked Fresh Top 10 returns 4.98\%, versus 6.45\% for Old targets and 13.87\% for Reset. At Top 30, the corresponding figures are 6.00\%, 5.01\% and 1.48\%. Including the evolving-model transition changes the Fresh/Old/Reset averages to 10.22\%/5.01\%/15.67\% at Top 10 and 7.06\%/3.62\%/4.39\% at Top 30. The reversal is visible at both levels of aggregation: which stocks enter the shortlist matters more than how close the methods are in average forecast error.

The common 203-stock panel is used for the accuracy comparison. Computing accuracy only on each method's own Top 10 would change both the forecasts and the evaluated stocks at once. The selected-stock return tables instead answer the investment question directly: what happened to the stocks that each ranking chose? A same-union accuracy diagnostic is retained in the artifacts for readers tracing the selection boundary.

\subsection{Six-Month Common-Cutoff Forecast Sensitivity (\texorpdfstring{$3\to4$}{3 to 4})}

The longer check follows the first refresh because its second-quarter horizon has nearly matured by the frozen market cutoff. The later transitions' second-quarter targets fall on 4 December 2026 and 2--3 January 2027. The first transition therefore supplies the available longer-horizon comparison, while the short-window tests retain all three transitions.

At the 30 October 2026 market cutoff, the first transition ($3\to4$) has 177--179 calendar days of post-entry history across all 203 stocks (52 stocks have nominal Q2 targets on 30 October and 151 on 1 November). Evaluated at the common 30 October close, six-month MAE is 21.635 pp (RMSE 38.587 pp) for Old fixed targets, 20.495 pp (RMSE 37.288 pp) for Old paths reset, and 20.135 pp (RMSE 36.440 pp) for Fresh reports, yielding a Fresh-minus-Reset difference of -0.360 pp (95\% CI [-1.161, +0.457] pp). Table~\ref{tab:longer} shows that the Top 10 vs. Top 30 reversal persists over this longer horizon.
\begin{table}[!htbp]\centering\small
\caption{Nearly six-month quote-currency price-return sensitivity for the first transition through 30 October 2026.}\label{tab:longer}
\resizebox{\linewidth}{!}{\begin{tabular}{lrrr}
\toprule
Selection rule & Old targets & Old paths reset & Fresh reports \\
\midrule
Top 10: six-month forecasts & -2.47\% & 8.10\% & -1.07\% \\
Top 30: six-month forecasts & 1.91\% & -1.32\% & 3.40\% \\
Top 10: original quarter picks & -8.25\% & 9.01\% & -1.09\% \\
Top 30: original quarter picks & 1.34\% & 2.15\% & 5.23\% \\
\bottomrule
\end{tabular}
}
\end{table}

Table~\ref{tab:longer} separates two ways to examine that longer window. Its first two rows select stocks using forecasts to the six-month cutoff; the last two keep the original quarter-ranked selections and extend their holding period. This separates changing the forecast horizon used for selection from simply holding the same stocks longer.

For six-month-ranked stocks, Fresh Top 10 returns $-1.07\%$, compared with $8.10\%$ for Reset and $-2.47\%$ for Old targets. At Top 30, Fresh returns $3.40\%$, Reset $-1.32\%$ and Old targets $1.91\%$. Keeping the original quarter-ranked stocks and extending their holding period preserves the same Fresh-versus-Reset direction at each breadth. The small difference in average forecast error thus sits alongside opposite investment rankings for ten-stock and thirty-stock selections.

\section[Interpolation, Realization Dates, and Cohort Checks]{Interpolation, Realization Dates,\\and Expanding-Cohort Checks}

Table~\ref{tab:sensitivity} verifies that our primary accuracy conclusions are invariant to (i) using linear percentage-return interpolation instead of geometric wealth interpolation between quarterly knots, and (ii) evaluating forecast curves at the exact saved realization close date $s_T \le T$ rather than nominal expiry $T$ when exchange holidays create 1-to-3 day gaps.
\begin{table}[!htbp]\centering\small
\caption{Fresh-minus-Reset MAE under alternative interpolation and realization-date conventions.}\label{tab:sensitivity}
\resizebox{\linewidth}{!}{\begin{tabular}{lrrr}
\toprule
Series & Convention & $\Delta$ MAE (pp) & 95\% interval \\
\midrule
Same model & Geometric (primary) & -0.040 & [-0.380, +0.289] \\
Same model & Linear interpolation & -0.042 & [-0.380, +0.286] \\
Same model & Realization-date expiry & -0.041 & [-0.379, +0.287] \\
Evolving & Geometric (primary) & -0.081 & [-0.386, +0.244] \\
Evolving & Linear interpolation & -0.082 & [-0.387, +0.242] \\
Evolving & Realization-date expiry & -0.081 & [-0.385, +0.242] \\
\bottomrule
\end{tabular}
}
\end{table}

Relaxing the balanced-panel restriction to include all split-clean stocks with complete 10-report councils in each individual transition expands coverage to 208, 268, and 271 stocks in $3\to4$, $4\to5$, and $5\to6$, respectively, yielding Fresh-minus-Reset $\Delta\text{MAE}$ values of -0.387 pp ([-0.915, +0.126]), +0.088 pp ([-0.519, +0.711]), and +0.381 pp ([-0.059, +0.831]; see revision\_series.json).

\section{Portfolio Breadth (Top 30), Endpoint, and Safeguard Sensitivities}

Table~\ref{tab:top30} reports the full Top 30 USD buy-and-hold portfolio backtest under the exact same clocks, 10 bps per-leg trading fees, and upfront compute cost deductions as Table~\ref{tab:portfolio}. Table~\ref{tab:robustness} reports the common 365-day forecast endpoint sensitivity and symmetric D3 shock-guard results across both Top 10 and Top 30.
\begin{table}[!htbp]\centering\small
\caption{Top 30 buy-and-hold portfolio returns (USD) under the same fee and all-in operating-cost assumptions as Table~\ref{tab:portfolio}.}\label{tab:top30}
\resizebox{\linewidth}{!}{\begin{tabular}{lrrrrr}
\toprule
Window & Old fixed & Old + guard & Path reset & Fresh & SPY \\
\midrule
3$\to$4 (~3mo) & 6.75\% & 5.70\% & 1.62\% & 5.39\% & 1.36\% \\
4$\to$5 (~3mo) & 8.77\% & 7.61\% & 9.47\% & 11.44\% & 4.47\% \\
5$\to$6 (~3mo) & 5.48\% & 2.64\% & 2.41\% & -2.66\% & 1.97\% \\
Same-model mean & 6.12\% & 4.17\% & 2.01\% & 1.37\% & 1.66\% \\
3$\to$4 (~6mo) & 3.24\% & 5.71\% & -1.27\% & 16.86\% & 3.46\% \\
\bottomrule
\end{tabular}
}
\end{table}

\begin{table}[!htbp]\centering\small
\caption{Common 365-day endpoint and symmetric-safeguard portfolio returns after modeled costs.}\label{tab:robustness}
\resizebox{\linewidth}{!}{\begin{tabular}{lrrrr}
\toprule
Window & Top $k$ & Common: old + guard & Common: fresh & Common: both guarded \\
\midrule
3$\to$4 (~3mo) & 10 & 2.44\% & 4.55\% & 4.55\% \\
3$\to$4 (~3mo) & 30 & 5.13\% & 5.39\% & 4.60\% \\
4$\to$5 (~3mo) & 10 & -0.69\% & 22.45\% & 22.45\% \\
4$\to$5 (~3mo) & 30 & 7.46\% & 11.44\% & 11.44\% \\
5$\to$6 (~3mo) & 10 & 2.07\% & 3.93\% & 3.93\% \\
5$\to$6 (~3mo) & 30 & 2.64\% & -0.36\% & -0.36\% \\
3$\to$4 (~6mo) & 10 & -11.26\% & 8.55\% & 8.55\% \\
3$\to$4 (~6mo) & 30 & 5.06\% & 16.86\% & 15.91\% \\
\bottomrule
\end{tabular}
}
\end{table}

\section{Reference Token Cost Formula}

For report task $j$ with total input tokens $I_j$, recorded cached input tokens $K_j$ (where unlogged cache fields are set to $K_j = 0$), visible output tokens $O_j$, and reasoning/thinking tokens $H_j$, the normalized standard reference cost under the 2 November 2026 Gemini 3.1 Pro Preview schedule ($\le 200\text{k}$ prompt tier) is:

\begin{equation}
c_j = 10^{-6}\Big[2.00(I_j - K_j) + 0.20 K_j + 12.00(O_j + H_j)\Big]\ \mathrm{USD}.
\end{equation}
The batch-API reference illustration substitutes 1.00 and 6.00 for the uncached input and output multipliers while retaining 0.20 for cached input tokens. Summing across all 6,090 fresh report tasks (609 stock councils across Batches 4, 5, and 6) yields a total standard reference cost of USD 1,486.27 and a batch reference cost of USD 754.36.

\begin{samepage}
\section{Archive Provenance and Checksum Contract}

The frozen evaluation archive was compiled on 2026-11-02 04:31:25 UTC with market observations cut off at 2026-10-30. The canonical source archive SHA-256 digest is:

\begin{center}\footnotesize\ttfamily
8504a1b196fcf0ef8184061242cf20f9b983\\
ba79bf5a49c6fedc8098c61221e8
\end{center}

Version 2 changes the capital and operating-cost scenarios, including recurring budget hurdles; the frozen market cutoff and stock selections are unchanged. The exact wallet-engine and shock-guard SHA-256 digests are retained with the results in \texttt{portfolio\_backtest.json}.
\end{samepage}

\begingroup\small
\begingroup\small
\begin{thebibliography}{10}
\setlength{\itemsep}{0pt}\setlength{\parskip}{0pt}
\bibitem{buffett1996} Warren E. Buffett. \emph{Chairman's Letter, 1996}. Berkshire Hathaway. \url{https://www.berkshirehathaway.com/letters/1996.html}.
\bibitem{frugalgpt} Lingjiao Chen, Matei Zaharia, and James Zou. \emph{FrugalGPT: How to Use Large Language Models While Reducing Cost and Improving Performance}. 2023. \href{https://arxiv.org/abs/2305.05176}{arXiv:2305.05176}.
\bibitem{forecastbench} Ezra Karger, Houtan Bastani, Chen Yueh-Han, Zachary Jacobs, Danny Halawi, Fred Zhang, and Philip E. Tetlock. \emph{ForecastBench: A Dynamic Benchmark of AI Forecasting Capabilities}. ICLR, 2025. \href{https://arxiv.org/abs/2409.19839}{arXiv:2409.19839}.
\bibitem{lastmile} Yuhua Liao, Zetian Wang, Qiangqiang Nie, and Zhenhua Zhang. \emph{Bridging the Last Mile of Time Series Forecasting with LLM Agents}. 2026. \href{https://arxiv.org/abs/2606.02497}{arXiv:2606.02497}.
\bibitem{simplicity} Meiru Zhang, Auss Abbood, Zaiqiao Meng, and Nigel Collier. \emph{The Power of Simplicity in LLM-Based Event Forecasting}. Proceedings of REALM 2025, pp. 454--470. \href{https://doi.org/10.18653/v1/2025.realm-1.32}{doi:10.18653/v1/2025.realm-1.32}.
\bibitem{ramdowar-agree} Russlan Ramdowar. \emph{When AI Advisors Agree: Distinguishing Forecast Risk from Forecast Reliability}. Version 2, 2026. \href{https://arxiv.org/abs/2610.13431v2}{arXiv:2610.13431v2}. Research artifacts: \href{https://doi.org/10.5281/zenodo.23373782}{doi:10.5281/zenodo.23373782}.
\bibitem{ramdowar-models} Russlan Ramdowar. \emph{When Models Judge Models: How Source Identity and Disagreement Shape Multi-Agent Financial Forecasts}. 2026. \href{https://doi.org/10.5281/zenodo.23083356}{doi:10.5281/zenodo.23083356}.
\bibitem{spo} Adam N. Elmachtoub and Paul Grigas. \emph{Smart ``Predict, then Optimize''}. Management Science, 68(1):9--26, 2022 (published online 2021). \href{https://doi.org/10.1287/mnsc.2020.3922}{doi:10.1287/mnsc.2020.3922}.
\bibitem{backtest} David H. Bailey, Jonathan M. Borwein, Marcos L\'opez de Prado, and Qiji Jim Zhu. \emph{The Probability of Backtest Overfitting}. Journal of Computational Finance, 20(4), 2017. \href{https://doi.org/10.21314/JCF.2016.322}{doi:10.21314/JCF.2016.322}.
\bibitem{pricing} Google. \emph{Gemini Developer API Pricing: Gemini 3.1 Pro Preview}. Official price schedule, accessed 2 November 2026. \url{https://ai.google.dev/gemini-api/docs/pricing}.
\end{thebibliography}
\endgroup
\endgroup
\end{document}